\documentclass[11pt]{article}
\usepackage[utf8]{inputenc}
\usepackage{amsmath, amssymb, graphicx, natbib, authblk}
\usepackage{geometry}
\usepackage{amsmath}
\usepackage{color}
\usepackage{amssymb}
\usepackage{psfrag}
\usepackage{listings}
\usepackage{array}
\usepackage[hyphens]{url}
\usepackage[hidelinks]{hyperref}
\hypersetup{breaklinks=true}
\usepackage{algorithm}
\usepackage[noend]{algpseudocode}

\newcommand\Rey{\mbox{\textit{Re}}}  % Reynolds number

\newcommand{\vecx}{\mathbf{x}}
\newcommand{\vecu}{\mathbf{u}}
\newcommand{\vecq}{\mathbf{q}}

\newcommand{\vecf}{\mathbf{f}}
\newcommand{\vecA}{\mathbf{\widehat{\delta^l_\xi x \,}}}
\newcommand{\vecB}{\mathbf{\widehat{\delta^m_\eta x \,}}}
\newcommand{\vecC}{\mathbf{\widehat{\delta^n_\zeta x \,}}}
\newcommand{\N}{\mathcal{N}}
\newcommand{\A}{\mathcal{A}}

\newcolumntype{C}[1]{>{\centering\let\newline\\\arraybackslash\hspace{0pt}}m{#1}}

\title{A Chimera method for high-fidelity simulation of turbulent flows}

\author[1]{Andrea Di Mascio\thanks{Email: \texttt{andrea.dimascio@univaq.it}}}
\author[2]{Sergio Pirozzoli\thanks{Email: \texttt{sergio.pirozzoli@uniroma1.it}}}

\affil[1]{Dipartimento di Ingegneria Industriale e dell'Informazione e di Economia, Università degli Studi dell'Aquila, Piazzale Ernesto Pontieri, Monteluco di Roio, 67100 L’Aquila, Italy}

\affil[2]{Dipartimento di Ingegneria Meccanica e Aerospaziale, Sapienza Università di Roma, Via Eudossiana 18, 00184 Rome, Italy}

\date{\today}

\begin{document}

\maketitle

\begin{abstract}
We develop a block-structured solver for high-fidelity simulation of flows in complex geometries, 
based on overlapping (Chimera) meshes. The key components of the algorithm are a 
baseline dissipation-free central discretization and selective high-order filtering, 
which ensure uniform accuracy and minimal numerical diffusion. These favorable 
properties are preserved through efficient interpolation across overlapping blocks.
Numerical tests demonstrate that the method guarantees a uniform order of accuracy 
even for distorted, overlapping meshes. Tests conducted for turbulent flow in a pipe show 
no significant issues at the interfaces or overlapping blocks, highlighting the method's 
potential for direct numerical simulation (DNS). Finally, we show that the method 
performs satisfactorily for geometrically complex problems, such as flow past a ship propeller, 
where it accurately replicates experimental data. 
\end{abstract}

% \linenumbers


\section{Introduction}
\label{sec:introduction}
Direct numerical simulation of turbulence (DNS) started several decades
ago as a pioneering tool for the exploration of fundamental physical
phenomena. Early studies~\citep{orszag_72,kim_87} were limited to
extremely simple geometries with periodic boundary conditions along
two of three spatial directions, and to low Reynolds numbers
corresponding to barely turbulent flow conditions. Over the years,
DNS has mainly remained a tool to analyze relatively simple flows
including homogeneous directions, although the accessible Reynolds
numbers have greatly increased. As a recent example, turbulent flow
in a circular pipe up to bulk Reynolds number ($\Rey_b = D u_b / \nu$,
where $u_b$ is the mean flow velocity, $D$ is the pipe diameter,
and $\nu$ is the fluid kinematic viscosity) $\Rey_b=615,000$ has been
reached, corresponding to typical operating conditions in oil
pipelines~\citep{pirozzoli_24}.

The fast growth in accessible Reynolds numbers has been made possible
by the exponential growth in the computing power of supercomputers,
following Moore's law with good precision
(\url{https://www.top500.com}). Progress on the side of geometrical
complexity has not been as rapid, although flows of technological
interest are starting to be approached~\citep{hosseini_16}. This
shortcoming is mainly due to the fact that DNS requires
computationally efficient algorithms, which are typically available
only for simplified geometrical settings. Typical DNS algorithms rely
on the use of Cartesian or curvilinear structured grids, which allow
for high accuracy with excellent computational efficiency. However,
flows with even moderate geometrical complexity cannot be easily
accommodated on structured mesh systems.

The use of unstructured meshes is the natural alternative; however,
computational efficiency here is severely hampered by memory stride
related to the difficulty of efficiently storing arrays in the
absence of natural data ordering. A possible alternative to the use of
unstructured meshes involves the so-called Chimera approach, first
introduced for the Euler equations by \citet{benek_83}. In essence,
the method involves the use of one (or more) body-fitted grids overlaid
on a background (typically Cartesian) mesh. The separate generation of
a body-fitted grid and a background grid allows great simplifications
in the meshing process and at the same time allows modeling of complex
systems with relative motion of parts of a structure, e.g., helicopter
rotors~\citep{pomin_02}.
Previous attempts to carry out DNS using an unstructured (non-overlapping)
mesh include the use of nested grids~\citep{kravchenko_96}, whereby
the mesh spacing in each coordinate direction is made to increase with
the wall distance. Issues associated with proper data exchange at the
interface between zones with different spacing were pinpointed in that
study, which reflected in imperfect statistics of velocity fluctuations.

The goal of this paper is to explore the capabilities of a novel Chimera
algorithm to carry out high-fidelity numerical simulations of turbulent flow,
including DNS, in an accurate and efficient fashion. 
The structure of the paper is the following: in section~\ref{sec:numerics}
we describe the numerical strategy, with particular emphasis on accurate
and stable data exchange between the structured mesh and the underlying
Cartesian mesh; in section~\ref{sec:results}, we present results of
numerical tests to establish the formal accuracy of the method on
simple benchmark problems, and then we apply it to the DNS of flow in
a circular pipe in section~\ref{sec:pipeflow}. Finally, we utilize the
described algorithm in the Detached Eddy simulation of the flow past
a marine propeller in section~\ref{sec:propeller}, illustrating its
applicability to engineering challenging problems. Concluding remarks
wrap up the paper in section~\ref{sec:conclusions}.

\section{Numerical method}
\label{sec:numerics}

In this section, we outline the numerical algorithm, which is based on
high-order central finite-difference discretization, with reference to
block-structured grids with partial overlap.
This approach is particularly well-suited for high-fidelity simulations 
of turbulent flow as it guarantees minimal, or even zero, numerical dissipation.
First, we discuss high-order central schemes in curvilinear coordinate
systems.  We then provide details on evaluating the metric tensor in
three-dimensional curvilinear grids, including extensions to Chimera
grids.  Additionally, we address the application of high-order numerical
dissipation to control unphysical oscillations.  In the latter part of
the section, we explain grid topology computation, emphasizing techniques
for cutting holes in the mesh and connecting different structured blocks.

\subsection{Discrete model}

We consider a set of curvilinear coordinates defined by the mapping
$x_k = x_k (\xi_j)$, with $j,k=1,\ldots,d$ 
($d$ is the number of spatial dimensions), where $x_k$ expresses
positions in the physical space, and $\xi_j$ positions in the 
logical space.
We start with the Euler equations of fluid dynamics 
cast in the strong conservation form~\citep{vinokur_74},
\begin{equation}
	\frac{\partial (J {\bf q})}{\partial t} + \frac{\partial}{\partial \xi_k} \left( J \xi_{k,j} {\bf f}_j \right) = 0,
\end{equation}
with
\begin{equation}
 {\bf q} =
\left[
\begin{array}{c}
\rho \cr \rho u_i \cr \rho E
\end{array}
\right], \quad
{\bf f}_j =
\left[
\begin{array}{c}
\rho {u}_j \cr \rho u_i {u}_j + p \delta_{ij} \cr \rho {u}_j H
\end{array}
\right], \quad i=1,\ldots,d ,
\label{eq:euleqn}
\end{equation}
where $\rho$ is the local fluid density, $u_i$ denotes the Cartesian velocity component in the $i$-th direction, 
$E=p / \rho/ (\gamma-1) + u_i u_i/2$,
$H=E+p/\rho$,
and $J$ the determinant of the coordinate transformation.
Here and throughout the paper the repeated index convention
is used to indicate summation. For the present purposes 
we assume that the coordinate transformation is
locally invertible and sufficiently smooth.

\citet{pirozzoli2011stabilized} showed that Eqn.~\eqref{eq:euleqn}
can be approximated with order of accuracy $2L$ through  centred formulas like
\begin{equation}
D_j \vecf_{\N} = \sum_{\ell=1}^{L} a_{\ell} (\vecf_{j;\ell} - \vecf_{j;-\ell})_{\N} ,
\label{fin_diff}
\end{equation}
where
\begin{equation}
\left( \vecf_{j;\ell} \right)_\N = \vecf_{\xi_1,\xi_2,\ldots,\xi_{j+\ell},\ldots,\xi_d} .
\end{equation}
\citet{pirozzoli2011stabilized} 
proved that several spitting of the convective terms yield numerical stability in curvilinear coordinates.
In particular, he considered a locally conservative approximation of \eqref{eq:euleqn},
\begin{equation}
J \frac{d \vecq}{dt}+ \sum_{j=1}^d \left( \frac{\vecf_{j,\ell}-\vecf_{j,-\ell}} {h_j} \right)_\N = 0,
\label{vect_eq}
\end{equation}
with numerical flux defined as
\begin{equation}
\vecf_{j,\ell} =  2 \sum_{\ell=1}^L a_\ell  \sum_{m=1}^{\ell-1} 
\left[ 
\begin{array}{c}
\widetilde{ \left( \rho, \hat{u}_j , 1 \right)}_{j;-m,\ell} 
\\*[3mm]
\widetilde{\left( \rho , \hat{u}_j , \vecu \right)}_{j;-m,\ell} + 
\widetilde{\left( p , J \nabla \xi_j,1 \right)}_{j;-m,\ell}
\\*[3mm]
\widetilde{\left( \rho , \hat{u}_j , H \right)}_{j;-m,\ell}
\end{array} 
\right]_\N ,
\end{equation}
where 
\begin{equation}
\hat{u}_j = J \vecu \cdot \nabla \xi_j,
\end{equation}
and the symbol $\widetilde{ \left( f,g,h \right)}_{j;-m,\ell}$ is the two-point, 
three-variable discrete averaging operator in the jth direction
\begin{equation}
\widetilde{ \left( f,g,h \right)}_{j;-m,\ell} = \frac{1}{8}(f_{j;l}+f_{j;l-m})(g_{j;l}+g_{j;l-m})(h_{j;l}+h_{j;l-m}) .
\end{equation}
In Eqn.~\eqref{vect_eq}, $h_j$ denotes the spacing along the $j$-th direction
in the logical space. With no loss of generality, it is assumed to be unity
throughout the rest of the paper.

In the formulas presented in this section, we frequently need to evaluate metric terms such as \( J \nabla \xi_j \). The method for calculating the metric tensor with the desired level of accuracy, while maintaining a uniform flow on arbitrary overlapping meshes, is detailed in \ref{sec:preservation} and \ref{sec:metric_on_chimera}.

\subsection{High-order artificial dissipation}

The approximation of equations~\eqref{eq:euleqn} using central
difference schemes and a co-located variable arrangement can
result in odd-even decoupling of the numerical solution. This issue
is exacerbated on Chimera grids due to the interpolation procedures
at the boundaries, as discussed below.

To mitigate these high-frequency oscillations without compromising
the formal accuracy of order $2L$, we introduce artificial dissipation
terms. Assuming $L \geq 2$, we evaluate the divided differences
recursively up to order $2(L-1)$ in each spatial direction and for
each scalar component of $\vecq$, such as $u$,
\begin{equation}
\left\{
\begin{array}{l@{\hspace*{7mm}}l@{\hspace*{7mm}}l}
\Delta^1_i u = u_i-u_{i-1}, & i=2-L,\ldots,N+L &   \\*[5mm]
\Delta^{p}_i u  = \Delta^{p-1}_i u-\Delta^{p-1}_i u, & i=1+p-L,N+L, & p=2,\ldots, 2L-2
\end{array}
\right.
\label{findiff}
\end{equation}
Then, for each variable $u$, at each interface $i+1/2$, we evaluate the oscillation indicator
\begin{equation}
\varphi_u |_{i+1/2} = \frac{ \Delta^{(2L-2)}_{i+L} u } 
                 {( \left|\Delta^{(2L-3)}_{i+L} u\right|+\left|\Delta^{(2L-3)}_{i+L-1} u\right|) }
            \, \,
            \frac{  \Delta^{(2L-2)}_{i+L-1} u } 
                 { ( \left|\Delta^{(2L-3)}_{i+L-1} u\right|+\left|\Delta^{(2L-3)}_{i+L-2} u\right|) } .
\label{factorphi}
\end{equation}
Taking into account \eqref{findiff}, inspection of equation \eqref{factorphi} reveals that
\begin{equation}
-1 \le \varphi_u |_{i+1/2} \le 1,
\end{equation}
and the extreme values are attained only when both conditions
\begin{equation}
 \Delta^{(2L-3)}_{i+L} u \, \Delta^{(2L-3)}_{i+L-1} u < 0 ,
 \quad \mbox{and} \quad
 \Delta^{(2L-3)}_{i+L-1} u \, \Delta^{(2L-3)}_{i+L-2} u < 0,
\label{cond_osc}
\end{equation}
hold, i.e. when a high-frequency oscillation on $\Delta^{(2L-3)} u $ appears around the interface. 
Note that, when one of the condition in \eqref{cond_osc} holds, the corresponding ratio in \eqref{factorphi} is 
equal to $\pm 1$. For instance, if the second condition holds,
from the definition in \eqref{findiff}, one has
\begin{eqnarray*}
            \frac{  \Delta^{(2L-2)}_{i+L-1} u } 
                 { ( \left|\Delta^{(2L-3)}_{i+L-1} u\right|+\left|\Delta^{(2L-3)}_{i+L-2} u\right|) }  & = & 
             =  \frac{ \Delta^{(2L-3)}_{i+L-1} - \Delta^{(2L-3)}_{i+L-2}} 
                { ( \left|\Delta^{(2L-3)}_{i+L-1} u\right|+\left|\Delta^{(2L-3)}_{i+L-2} u\right|) } =  \\
           & = & \pm 
           \frac{ ( \left|\Delta^{(2L-3)}_{i+L-1} u\right|+\left|\Delta^{(2L-3)}_{i+L-2} u\right|) }
                { ( \left|\Delta^{(2L-3)}_{i+L-1} u\right|+\left|\Delta^{(2L-3)}_{i+L-2} u\right|) } = \pm 1 .
\end{eqnarray*}
Under this condition, Taylor series expansion shows that 
\begin{equation}
 \varphi_u |_{i+1/2} =
                \left[\frac{ u^{(2L-2)}_{\xi } }
                            { 2  |u^{(2L-3)}_{\xi}|} \right]_{i+1} h
                      + O(h^2) =
                \left[\frac{ u^{(2L-2)}_{\xi } }
                            { 2  |u^{(2L-3)}_{\xi}|} \right]_{i+1/2} h
                      + O(h^2) ,
\end{equation}
where $u^{(L')} = \partial^{L'} u / \partial \xi^{L'}$.
At all other interfaces 
\begin{equation}
\begin{array}{lcl}
 \varphi_u |_{i+1/2} 
& = & \displaystyle
\frac{  [u^{(2L-2)}_{\xi} ]_{i+1}}
     {2 |u^{(2L-3)}_{\xi}|_{i+1} } 
\frac{  [u^{(2L-2)}_{\xi} ]_{i}}
     {2 |u^{(2L-3)}_{\xi}|_{i} } h^2 + O(h^4)
\\*[5mm]
& = & \displaystyle
                \left[\frac{ u^{(2L-2)}_{\xi } }
                            { 2  |u^{(2L-3)}_{\xi}|} \right]^2_{i+1/2} h^2
                      + O(h^4) .
\end{array}
\end{equation}
We then introduce the coefficient
\begin{equation}
\tilde{\varphi} |_{i+1/2} = \max (0,-\varphi |_{i+1/2}),
\end{equation}
which is positive only where $[u^{(2L-2)}_{\xi} ]_{i+1} \, \, [u^{(2L-2)}_{\xi} ]_{i} < 0$,
and which is unity
only where both conditions in \eqref{cond_osc} hold; elsewhere, it is $O(h)$.

Finally, we determine the artificial dissipative flux at the interface $i+1/2$ for each variable as 
\begin{equation}
\begin{array}{lll}
f^a_u |_{i+1/2} & = & 
\displaystyle \kappa \frac{(-1)^{L}}{2^{2L-1}} \,\, 
\lambda_{\max} |_{i+1/2} \,\,  |J \nabla \xi|_{i+1/2} \,\,
\tilde{\varphi} |_{i+1/2} \, 
(\Delta^{2L-2}_{i+L} u-\Delta^{2L-2}_{i+L-1} u)
\\*[5mm]
& = & 
\displaystyle
\A_{i+1/2} \tilde{\varphi} |_{i+1/2} \, 
(\Delta^{2L-2}_{i+L} u-\Delta^{2L-2}_{i+L-1} u) ,
\end{array}
\label{diff_flux}
\end{equation}
where $\kappa = O(1)$, 
$\lambda_{\max}$ is the maximum eigenvalue of the Jacobian matrix of $\vecf_{j;\ell}$, namely
\begin{equation}
\lambda_{\max} |_{i+1/2} = \max( \lambda_{\max} |_{i} , \lambda_{\max} |_{i+1}),
\end{equation}
and finally
\begin{equation}
\A_{i+1/2} = \kappa \frac{(-1)^{L}}{2^{2L-1}}\, \lambda_{\max} |_{i+1/2} |J \nabla \xi|_{i+1/2}.
\end{equation}
Taylor series expansion proves that
\begin{equation}
\begin{array}{lll}
f^a_u |_{i+1/2} & =  & 
\displaystyle
\A_{i+1/2} 
       \left[\left(\frac{ u^{(2L-2)}_{\xi } \, h }
                  { 2  |u^{(2L-3)}_{\xi}|} \right)^\sigma  
       u^{(2L-1)}_{\xi } h^{2L-1}  \right]_{i+1/2} 
+ O(h^{2L+\sigma})
\\*[5mm]
& = & \displaystyle
\A_{i+1/2} 
       \left[\left(\frac{ u^{(2L-2)}_{\xi }}
                  { 2  |u^{(2L-3)}_{\xi}|} \right)^\sigma  
       u^{(2L-1)}_{\xi }  \right]_{i+1/2}  h^{2L+\sigma-1}
+ O(h^{2L+\sigma}) ,
\end{array}
\label{diff_flux_ord}
\end{equation}
where $\sigma=0,1$ or $2$. Note that $\sigma=0$ only at the interfaces where the sign of the 
order $2L-2$ derivative changes, in which case $f^a_u = O(h^{2L-1})$. 
At all the other points $f^a_u = O(h^{2L})$ at least (it is exactly zero at interfaces where
$[u^{(2L-2)}_{\xi} ]_{i+1} \, \, [u^{(2L-2)}_{\xi} ]_{i} > 0$), and therefore the formal 
order of accuracy of the original central scheme is retained.

Linear stability analysis and numerical tests convey that the time step 
restriction due to this form of artificial dissipation is
\begin{equation}
\displaystyle
\Delta t  \le \min_{i,\, \tilde{\varphi} > 0} \, \, \frac{C}{\kappa \left(
\left[  \tilde{\varphi} \, |\nabla \xi| \, \lambda_{\max} \right]_{i-1/2}
+
\left[  \tilde{\varphi} \, |\nabla \xi| \, \lambda_{\max} \right]_{i+1/2}
\right)} ,
\end{equation}
where $C=O(1)$ depends on the time integration scheme.
For all the test cases herein reported, we have used $\kappa = 0.1$; 
hence, the incurred time step constraint is very mild.
When second-order discretization is used ($L=1$), 
we use the same artificial dissipation as for $L=2$

\subsection{Grid topology}
\label{sec:topology}

In the following, we assume that the mesh consists of blocks $\mathcal{B}_b$, $b=1,2,\ldots,N_B$, each
with structured ordering and including $n^b_i \times n^b_j \times n^b_k$ points in directions $i$, $j$ and $k$, respectively.
The blocks can have partial overlap.
The computation of the flow on Chimera grids requires the knowledge
of active cells and holes in every block, together with block-to-block
connection to interpolate the solution where needed.  The rationale for
identifying active cells and holes in the grid is the most obvious,
as we want to retain the best discretization available at each point.
For this purpose, we must check each cell to verify whether
\begin{enumerate}
\item there is another cell overlapping with the candidate cell;
\item the overlapping cell is "smaller" in the sense specified in the following.
\end{enumerate}
A similar criterion guides the selection of donors for the
interpolation of the solution among blocks at interior points
and at those block interfaces that are neither natural
boundaries (e.g. walls, symmetry planes, inflow, outflow) nor matching
faces between blocks (i.e. where there is an actual continuity in the grid).
In both cases, we may find more than one possible donor. In this
situation, we choose the donor whose size is closest to the
cell under analysis.
As for the definition of the cell size ($d$), one can adopt either $d=\sqrt[3]{V}$ ($V$ being the cell volume)
or the cell diagonal $d=d_c$. When dealing with high-Reynolds-number flows with highly stretched grids,
in order to retain all the cells close to solid walls, we define the cell dimension as
\begin{equation}
   d=\min(d_n,d_c)  \qquad \mbox{or} \qquad d=\min(d_n,\sqrt[3]{V}),
\end{equation}
where $d_n$ is the distance from the wall. With this definition, $d \rightarrow 0$ near the walls,
and no other cell can cancel those in the boundary layers.

\subsection{Search algorithm for donors}

\begin{figure}
\centering
	\includegraphics[width=\textwidth,clip]{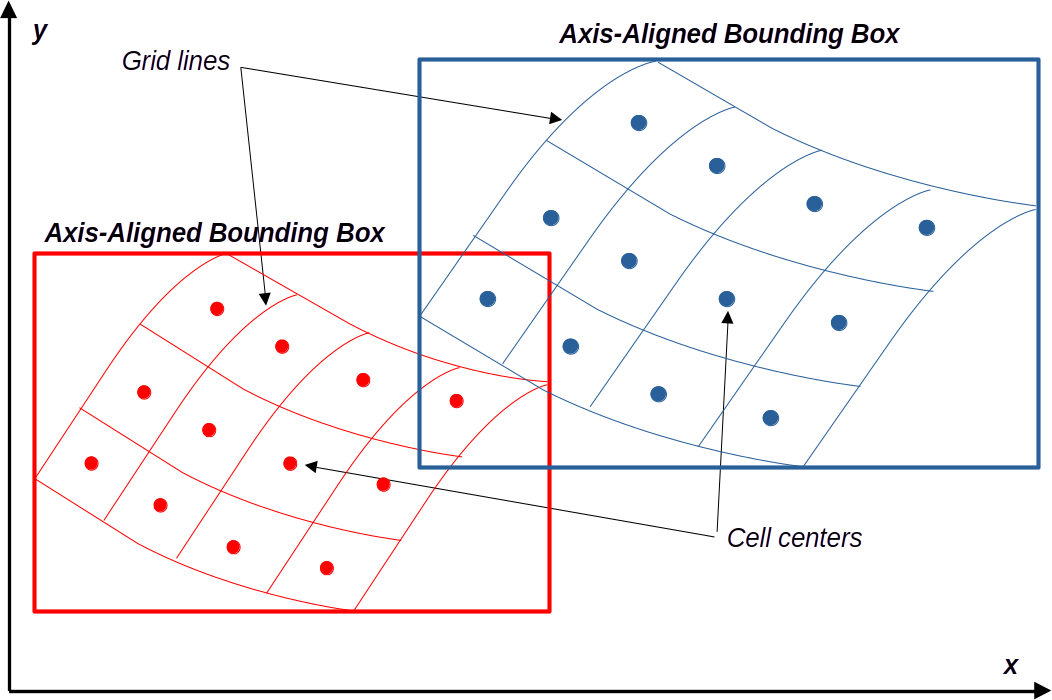}
\caption{Sketch of grid lines, cell centres and bounding boxes }
\label{fig:BoundingBox}
\end{figure}
\begin{figure}
\centering
	\includegraphics[width=\textwidth,clip]{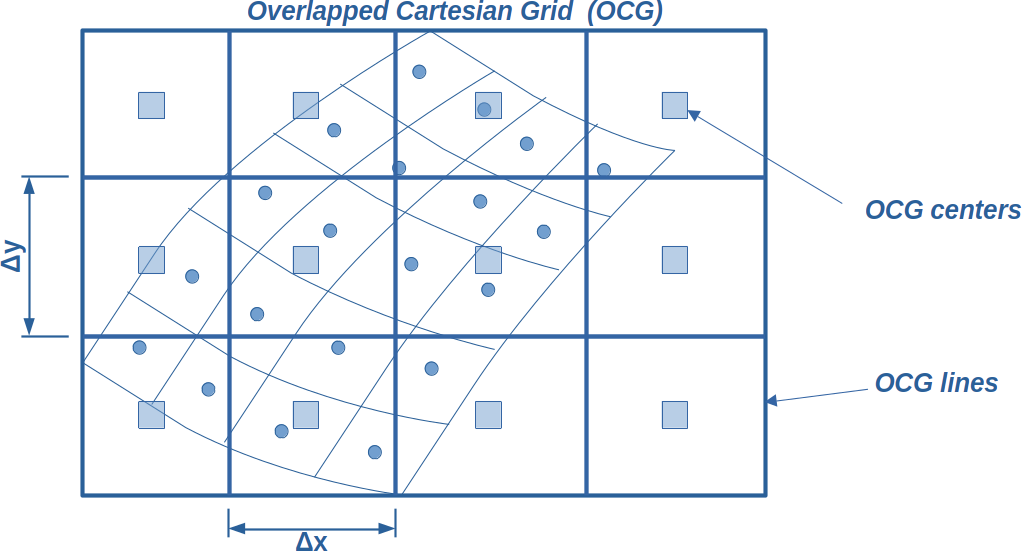}
\caption{Sketch of overlapping Cartesian grid }
\label{fig:OverlappedCartesianGrid}
\end{figure}

The computation of grid topology must be efficient to avoid negatively
impacting the overall CPU time, particularly in problems involving
moving boundaries~\citep{di2022vortex}, where the grid topology must
be updated at each time step. To accelerate the process of hole
identification and donor computation, we draw inspiration from the
classical Overlapped Cartesian Grid procedure developed by
\citet{verlet_67}, widely used in particle methods. This approach has
been adapted to work effectively with block-structured meshes that
allow for partial overlap.
The first step involves constructing an Axis-Aligned Bounding Box
(AABB) for each block $\mathcal{B}_b$. This box is defined as the
smallest axis-aligned parallelepiped that contains all the points
within the block. This method ensures computational efficiency while
maintaining compatibility with the block-structured grid framework.
This bounding box is such that
\begin{equation}
    \left\{
    \begin{array}{l}
    x_{\min}^{BB} \le x_{i,j,k} \le x_{\max}^{BB} \\*[0.3cm]
    y_{\min}^{BB} \le y_{i,j,k} \le y_{\max}^{BB} \\*[0.3cm]
    z_{\min}^{BB} \le z_{i,j,k} \le z_{\max}^{BB} .
    \end{array} \right.  \qquad \forall (i,j,k) \in \mathcal{B}_b,
\end{equation}
Therefore, entire blocks can be excluded from the search for donors 
as one of the following six conditions is satisfied for 
the generic cell centre ($\vec{x}_c$):
\begin{equation}
    \begin{array}{l@{\hspace*{2cm}}l}
    x_{c}  < x_{\min}^{BB}, &  x_{c} > x_{\max}^{BB}, \\*[0.3cm]
    y_{c}  < y_{\min}^{BB}, &  y_{c} > y_{\max}^{BB}, \\*[0.3cm]
    z_{c}  < z_{\min}^{BB}, &  z_{c} > z_{\max}^{BB}. 
    \end{array} 
\end{equation}
Moreover, possible holes in block $p$ must search for possible donors in block $q$ only when there is a partial
overlap between the corresponding bounding boxes, as sketched in figure~\ref{fig:BoundingBox}
for a two-dimensional problem.

Once the AABB for each block is identified, it is partitioned
into a Cartesian grid with uniform spacing in each direction, as
illustrated in Figure~\ref{fig:OverlappedCartesianGrid}. Generally,
the grid spacing satisfies $\Delta x \ne \Delta y$.
The number of points in the Overlapped Cartesian Grid (OCG) is typically
much smaller than the number of points in the underlying curvilinear
mesh. For example, the number of points in each direction of the OCG
is often $n_i^{OCG} = O(n^b_i/4)$ to $O(n^b_i/8)$. As a result, the
total number of cells in the OCG is reduced to approximately
$O(n_i \times n_j \times n_k / 64)$ to $O(n_i \times n_j \times n_k / 512)$.
This significant reduction ensures minimal overhead in computer memory usage.
In each cell of the OCG, we store the index of the closest point in the
underlying grid and flag the empty cells. To achieve this, all OCG
cells are initially initialized with an invalid index (e.g., a negative
value if grid numbering starts from zero or one) and a large value for
the current distance (e.g., $10^{10}$ if the typical distance is $O(1)$).
Next, we visit all cell centers of the underlying curvilinear mesh,
determining the corresponding indices of the OCG cell where they belong,
\begin{equation}
\begin{array}{l}
 i_{OCG} = \left\lfloor \frac{x_{i,j,k}-x_{\min}^{BB}}{\Delta x} \right\rfloor + 1,  \\*[3mm]
 j_{OCG} = \left\lfloor \frac{y_{i,j,k}-y_{\min}^{BB}}{\Delta y} \right\rfloor + 1,  \\*[3mm]
 k_{OCG} = \left\lfloor \frac{z_{i,j,k}-z_{\min}^{BB}}{\Delta z} \right\rfloor + 1,
\end{array}
\end{equation}
where $\lfloor \cdot \rfloor$ denotes the integer part and
the numbering of the OCG starts from one. We then store the index of the cell centres if the distance is smaller 
than the last saved value. An invalid index remains in the OCG cells not visited in the loop.

The OCG thus identified is useful for quick search of possible donors with the 
procedure described in the pseudo-code~\ref{alg:search_donors_serial}.
\begin{algorithm}
\caption{Donor search - pseudocode}\label{alg:search_donors_serial}
\begin{algorithmic}
\ForAll{$\mathcal{B}_b \in \mathcal{M}$}   \Comment{(Mesh $\mathcal{M}$ made of blocks $\mathcal{B}_b , \,\, b=1,2,\ldots,N_B$ )}
\State
\ForAll{$\vec{x}_c \in \mathcal{B}_b$}   
   \State \textit{$\mathbf{C_c}$ =  cell around $\vec{x}_c$}
   \State \textit{$d_c$ =  dimension of $\mathbf{C_c}$}
   \State
   \ForAll{$\mathcal{B}_p \, ; \, p \ne b$ \, \,}  
      \State  
      \If{$ \vec{x}_c \in \mathcal{AABB}_p $ }     \Comment{($\mathcal{AABB}_p$ 
                                                     is the AABB for block $p$)}
         \State \textit{Find $\mathcal{OCG}_p$ cell} \Comment{(
                                                     $\mathcal{OCG}_p$ is the
                                                     OCG for block $p$)} 
         \State \textit{$p_0$ : index saved in the $\mathcal{OCG}_p$ cell}
         \State 
         \If{ \textit{$p_0$ is a valid index} } 
             \State 
             \State \textit{Start from $\vec{x}_{p_0}$ }
             \State \textit{Find the closest cell center $\vec{x}_{p}$ (see subsection \ref{subsec:local_search})}
             \State \textit{$\mathbf{C_p}$ =  cell around $\vec{x}_p$}
             \State 
             \If {\textit{$\vec{x}_c \in \mathbf{C_p}$}}
                \State \textit{$d_p$ = dimension of  $\mathbf{C_p}$}
                \If { $d_p < d_c$ }
                   \State \textit{$\mathbf{C_c} \rightarrow \, $  Hole}
                   \If {\textit{Already found a donor } }
                       \State \textit{$\mathbf{C_{previous}}$ =  old donor cell}
                       \State \textit{$d_{previous}$ = dimension of $\mathbf{C_{previous}}$ }
                       \If { $|d_{p} - d_c | < |d_{previous} - d_c |$}
                          \State \textit{$\mathbf{C_p} \rightarrow \, $ Donor }
                       \EndIf
                   \Else
                       \State \textit{$\mathbf{C_p} \rightarrow \, $ Donor }
                   \EndIf
                \EndIf
             \EndIf
             \State 
         \EndIf
      \EndIf
      \State
   \EndFor
   \State
\EndFor
\EndFor
\end{algorithmic}
\end{algorithm}

\subsection{Local search algorithm}
\label{subsec:local_search}

\begin{figure}[h]
\centering
	\includegraphics[width=\textwidth,clip]{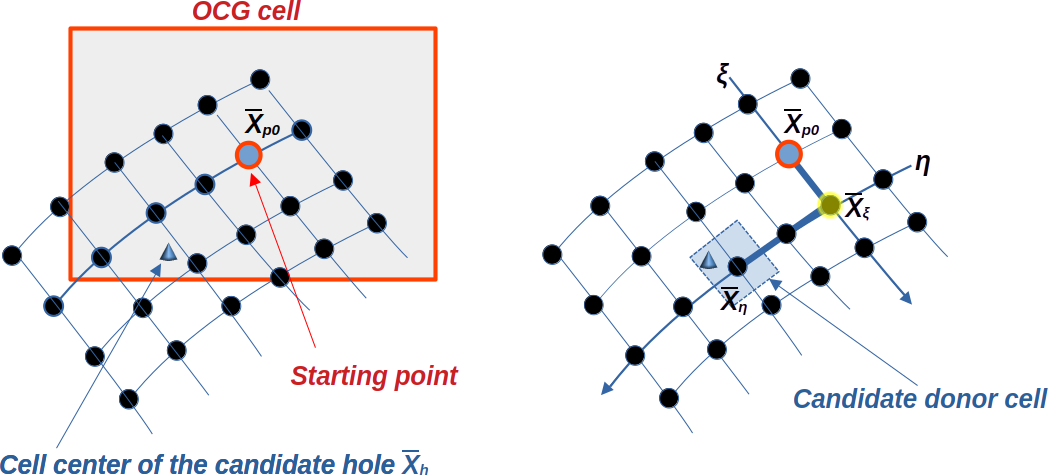}
\caption{Sketch of the search algorithm}
\label{fig:SearchAlgorithm}
\end{figure}

Next, we search for the closest cell center using the information stored
in the OCG, progressing along each curvilinear coordinate line.
Figure~\ref{fig:SearchAlgorithm} illustrates the local search algorithm
for a two-dimensional problem, with the three-dimensional case following
an analogous logic.
The search begins at point $\vec{x}_{p_0}$ in the figure. In each direction,
we evaluate the distance (specifically, the squared distance to save
CPU time) from the candidate hole to the two adjacent points and move
towards the closer one. For example, when moving along the $\xi$ coordinate
line in Figure~\ref{fig:SearchAlgorithm}, we find that point $\vec{x}_{\xi}$
is the closest to $\vec{x}_{h}$ along that line.
Next, we proceed along the other direction (denoted as $\eta$ in the
example) and repeat the procedure until we reach point $\vec{x}_{\eta}$,
which is the closest to $\vec{x}_{h}$ along the $\eta$ direction.
The search alternates between the $\xi$ and $\eta$ directions until it
converges at the point with the minimum distance, completing the process.
\begin{figure}[h]
	\includegraphics[width=\textwidth,clip]{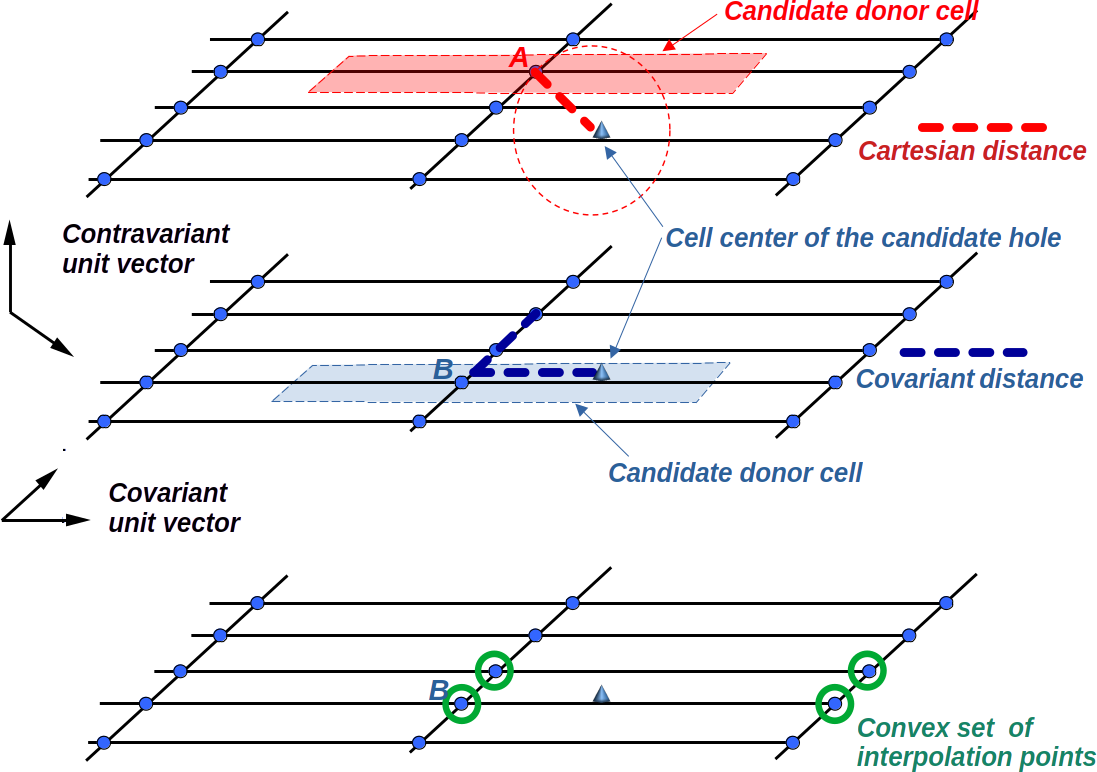}
\caption{Sketch of the search algorithm - Local search}
\label{fig:CovariantSearch}
\end{figure}

This simple search algorithm is efficient but can sometimes identify an
incorrect cell center as the candidate donor. This issue is illustrated
in the top panel of Figure~\ref{fig:CovariantSearch}: point $\mathbf{A}$
is the closest cell center of the block, but it does not lie within the
corresponding cell. This problem arises when the grid lacks high regularity.
To address this limitation for general grids, we refine the search by
checking a distance metric based on the covariant components of the
position vectors after initially identifying the closest point using
the original algorithm.
Specifically, any position vector can be expressed as
\begin{equation}
\vec{y} = \tilde{y}^i \vec{e}_i, \quad \text{where} \quad J \tilde{y}^i = \vec{y} \cdot \vec{e}^i,
\end{equation}
with $\vec{e}_i = \partial \vec{x} / \partial \xi_i$,
$\vec{e}^i = J \nabla \xi_i$, and $J$ representing the covariant vectors,
contravariant vectors, and the Jacobian of the coordinate transformation
$\vec{x}(\xi_k)$, respectively.
Using this formulation, we define the effective distance between two
points, $\vec{y}_A$ and $\vec{y}_C$, as:
\begin{equation}
\tilde{d}^2 = \left[ \tilde{y}_A^i - \tilde{y}_C^i \right]^2.
\end{equation}
Graphically, as shown in the middle panel of Figure~\ref{fig:CovariantSearch},
this distance is the sum of the squared lengths of the two segments
indicated by thick blue dashed lines.
By employing this definition of distance, the refined search algorithm
correctly identifies point $\mathbf{B}$ as the cell center containing
$\vec{x}_h$, resolving the issue encountered with point $\mathbf{A}$.

Once the candidate donor cell is identified, we accept it as a donor
if it meets the criteria outlined in the previous section. Finally,
we interpolate the solution from the active donor cells to the hole cells.
We have implemented two levels of interpolation accuracy:
\begin{itemize}
  \item \textit{second-order accurate interpolation}:
  In this method, we select a convex set of four cell centers in 2D
  (or eight in 3D), which includes the identified donor cell,
  surrounding the center of the hole. The solution is transferred
  from the donor and surrounding cells using bilinear interpolation
  in 2D (or trilinear in 3D). This process is illustrated in the bottom
  panel of Figure~\ref{fig:CovariantSearch}.
  \item \textit{fourth-order accurate interpolation}:
  For higher accuracy, we select a convex set of sixteen cell centers
  in 2D (or sixty-four in 3D), which also includes the identified
  donor cell, around the hole center. The solution is transferred
  from the donor and surrounding cells using the Cartesian product
  of two cubic interpolation functions in 2D (or three cubic functions
  in 3D). The set of points used for interpolation is defined by adding
  an additional layer of donor cells on each side of the set used
  for second-order interpolation.
\end{itemize}
These approaches ensure accurate solution transfer while accommodating
varying accuracy requirements.

\subsection{Forcing terms for the governing equations}

Once the grid topology is determined, we do not remove the holes from the discrete model.
Instead, we employ a technique adapted from the immersed-boundary method community~\citep{verzicco2023immersed}.
Specifically, we introduce a forcing term into the discretized equations at the hole locations, as follows:
\[
\frac{d \mathbf{q}_n}{d t}
+ \mathcal{L}_n^h(\mathbf{q}_{n-p}, \mathbf{q}_{n+q})
+ \frac{K}{\delta_n} (\mathbf{q}_n - \hat{\mathbf{q}}_n) = 0 \quad n = 1, \ldots, N \label{eq:forcing}
\]
where $\mathbf{q}_n$ is the state variable vector at point $n$,
$\mathcal{L}_n^h$ is the discrete operator at the same point,
$p$ and $q$ are parameters depending on the discrete model,
and $\hat{\mathbf{q}}_n$ is the interpolated state vector at point $n$.
Furthermore, $K$ is a constant of order $O(1) \div O(10)$,
and $\delta_n$ has the physical dimension of time, defined as:
\[
\delta_n = \min \left( \Delta t, \frac{\Delta \xi}{\lambda_\xi},
\frac{\Delta \eta}{\lambda_\eta}, \frac{\Delta \zeta}{\lambda_\zeta} \right),
\]
where $\Delta t$ is the time step, and $\Delta \xi, \Delta \eta, \Delta \zeta$ are the grid step sizes
in the coordinate directions $\xi, \eta, \zeta$, respectively.
The values $\lambda_\xi, \lambda_\eta, \lambda_\zeta$ represent the spectral radii of the Jacobian matrices
of the fluxes in each direction.
Several motivations suggest the use of this procedure:
\begin{enumerate}
      \item the data structure remains identical to traditional block-structured grids;
      \item all the algorithms developed for block-structured grids (e.g. geometrical multi-grid, alternate directions method, incomplete LU factorization) remain identical; hence, we can use existing software for structured blocks;
      \item when dealing with moving grids, we already have the values at re-activated cells without any 
            additional coding;
      \item the last term in equation \eqref{eq:forcing} causes exponential relaxation of the
            solution towards the interpolated value $\hat{\mathbf{q}}_n$ with time constant $\delta$:
            this term contributes to damping possible spurious waves forming between blocks with large differences in grid size.
\end{enumerate}

\section{Results}
\label{sec:results}

\subsection{Advection of homentropic vortex}

As an initial test, we considered the passive advection of a circular, homentropic, zero-circulation vortex, 
a benchmark case studied by many researchers. The initial velocity, density, and pressure are specified as follows:
\begin{eqnarray}
\frac{u(x,y)}{u_{\infty}} &=& 1 - \frac{M_v}{M_{\infty}} \, \frac{y-y_0}{r_v} \, e^{(1-\hat{r}^2)/2}, \nonumber \\
\frac{v(x,y)}{u_{\infty}} &=&     \frac{M_v}{M_{\infty}} \, \frac{x-x_0}{r_v} \, e^{(1-\hat{r}^2)/2}, \nonumber \\
\frac{\rho(x,y)}{\rho_{\infty}} &=& \left( 1 - \frac{\gamma-1}2 \, M_v^2 \, e^{1-\hat{r}^2} \right)^{\frac 1{\gamma-1}}, \nonumber \\ \label{eq:taylor}
\frac{p(x,y)}{p_{\infty}} &=& \left( 1 - \frac{\gamma-1}2 \, M_v^2 \, e^{1-\hat{r}^2} \right)^{\frac {\gamma}{\gamma-1}}, \nonumber 
\end{eqnarray}
where $(x_0, y_0)$ are the initial coordinates of the vortex center, $r_v$ is the radius of the 
vortex core, $M_v$ is the vortex Mach number (based on the maximum vortex-induced tangential velocity), 
$M_{\infty}$ is the free-stream Mach number, $\hat{r} = r/r_v$, and $r^2 = (x - x_0)^2 + (y - y_0)^2$. 
In our tests, we considered $M_v = 0.1$ and $M_{\infty} = 0.1$.
This test case is helpful because it has an analytical solution, as the
vortex is passively advection when the physical viscosity is
zero. The results of the numerical experiments are presented below for
both a block-structured Cartesian grid with overlap and a block-structured
curvilinear grid with overlap.

\begin{figure}[h]
\centering
	(a)   \includegraphics[width=6.00cm,clip]{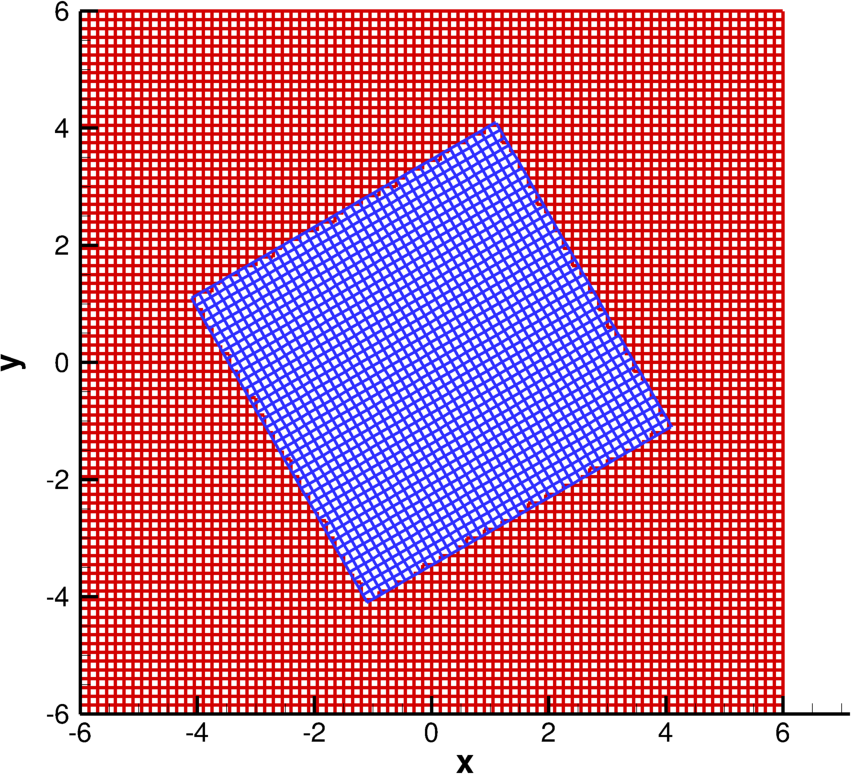}
	(b)   \includegraphics[width=6.00cm,clip]{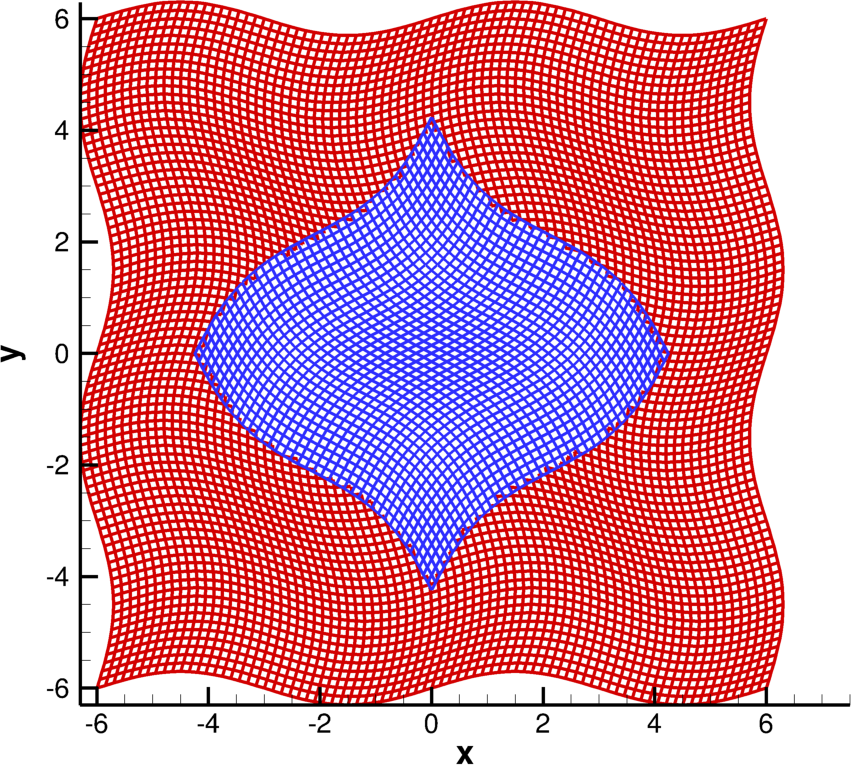}
\caption{Cartesian overlapping meshes (a) and curvilinear overlapping meshes (b) 
	for the homentropic vortex advection test case. 
	In both cases, the background mesh (red) includes $80 \times 80$ grid points, 
	and the internal mesh (blue) includes $40 \times 40$ grid points.}
\label{fig:vortex_mesh}
\end{figure}

\begin{figure}[h]
\centering
(a)   \includegraphics[width=6.00cm,clip]{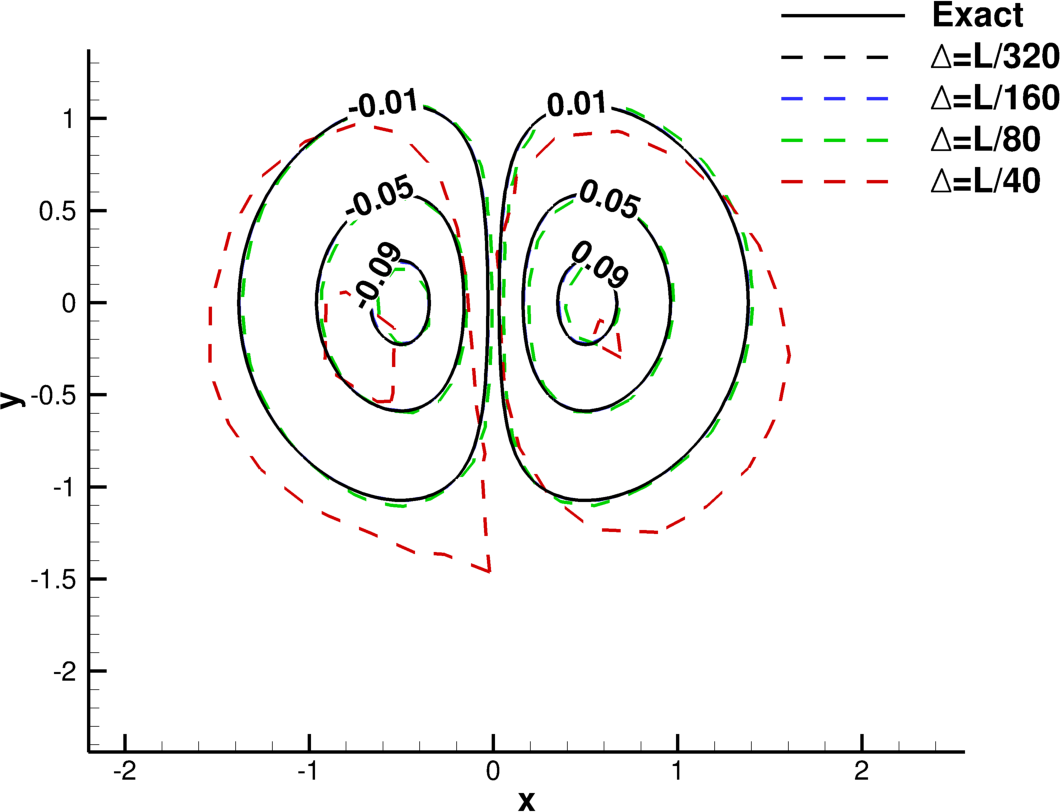}
(b)   \includegraphics[width=6.00cm,clip]{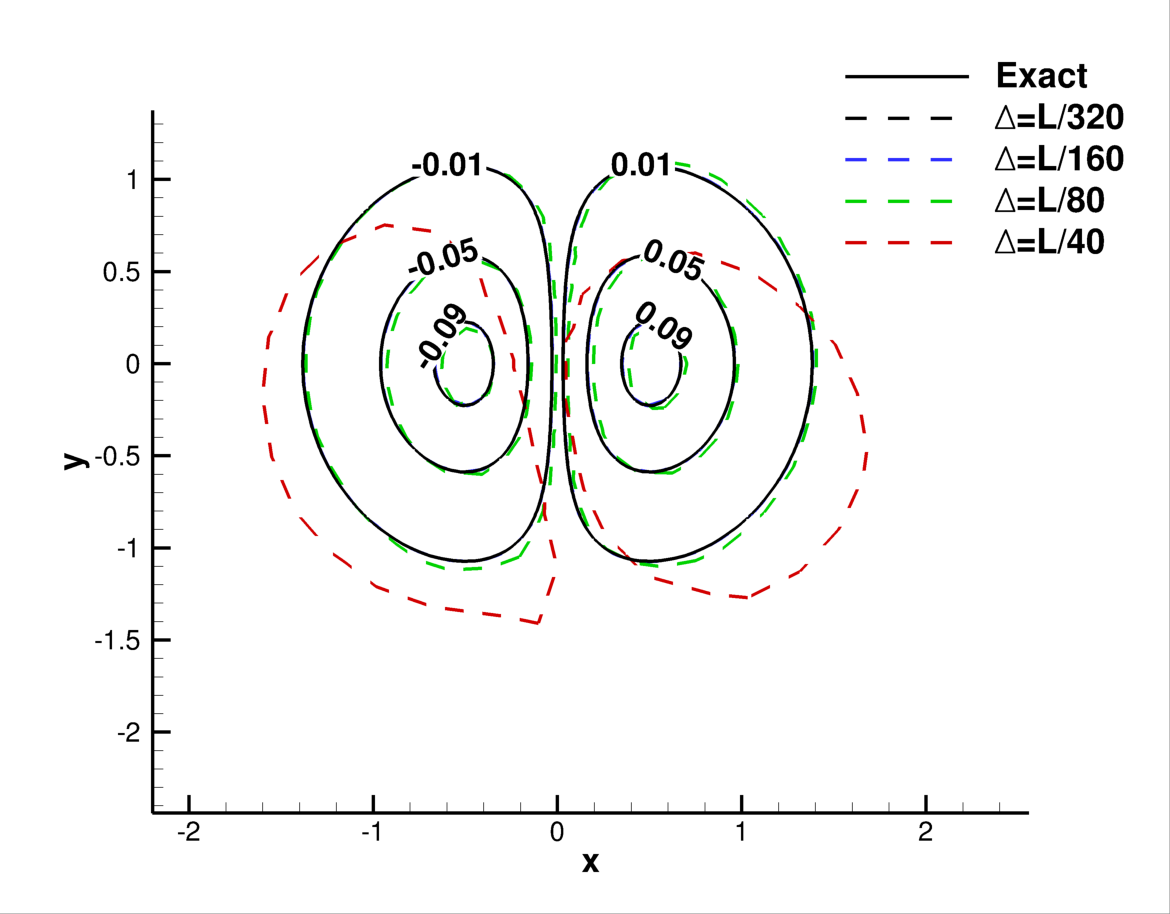}
\caption{Homentropic vortex advection: comparison of computed vertical velocity contours 
	with the exact solution (solid line), after one flow-through time. Results are shown for
	the Cartesian overlapping mesh (a) and the curvilinear overlapping mesh (b).}
\label{fig:vortex_cont}
\end{figure}

\begin{figure}
\centering
(a)   \includegraphics[width=6.00cm,clip]{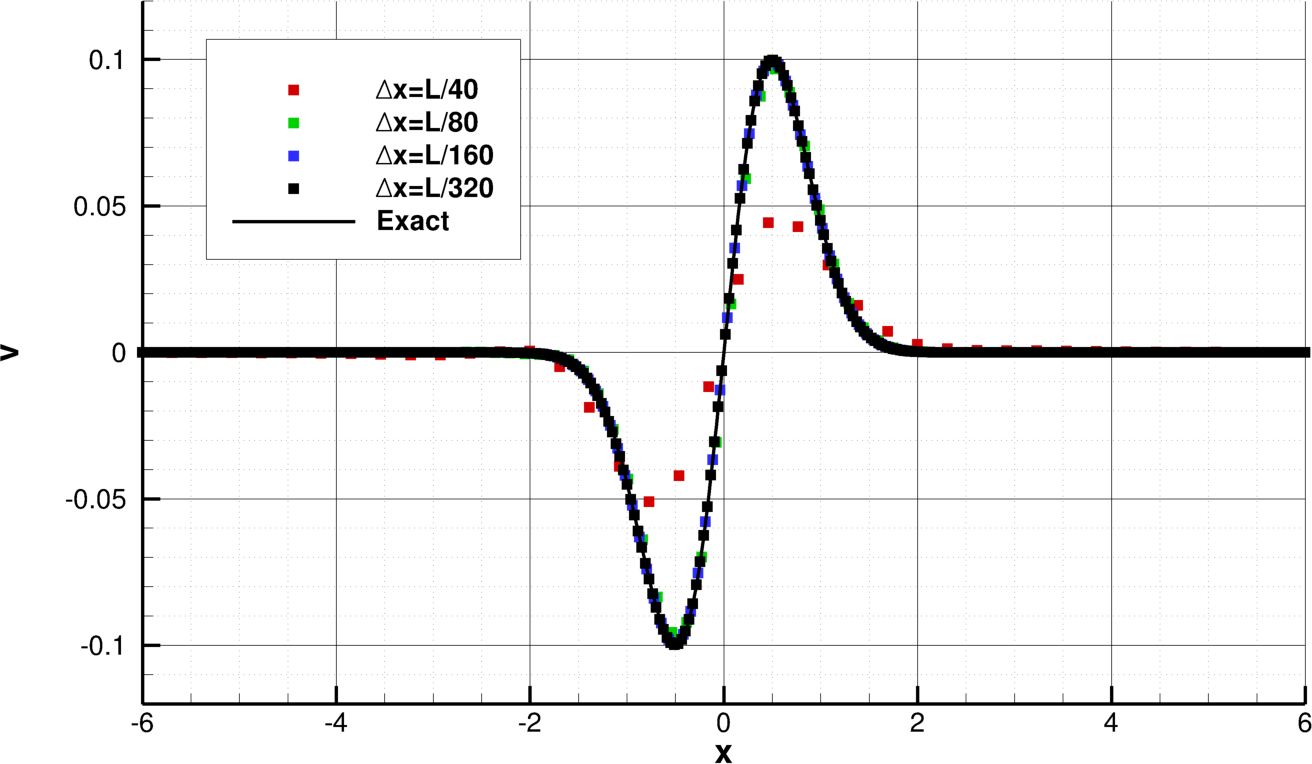}
(b)   \includegraphics[width=6.00cm,clip]{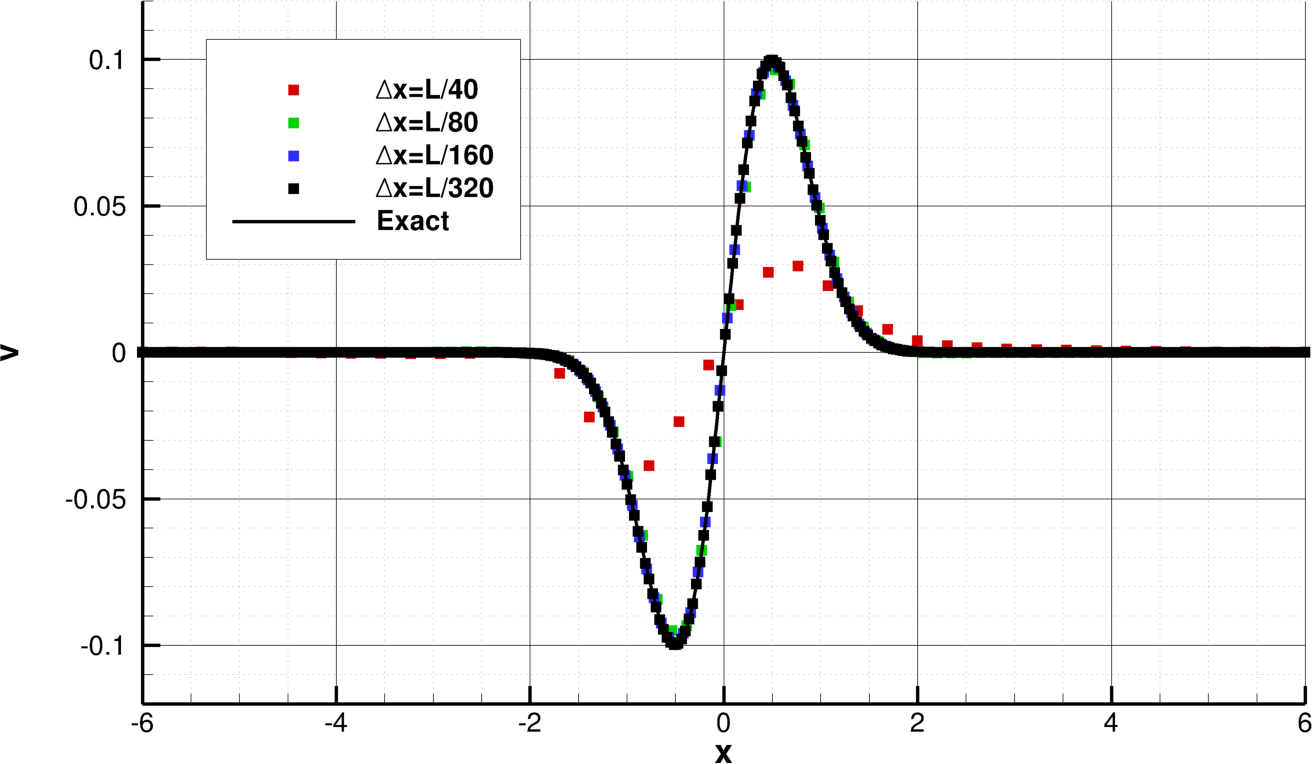}
\caption{Homentropic vortex advection: comparison of computed vertical velocity profiles
	with the exact solution (solid line), after one flow-through time. Results are shown for
	the Cartesian overlapping mesh (a) and the curvilinear overlapping mesh (b).}
\label{fig:vortex_prof}
\end{figure}

\begin{figure}
\centering
(a)   \includegraphics[width=6.00cm,clip]{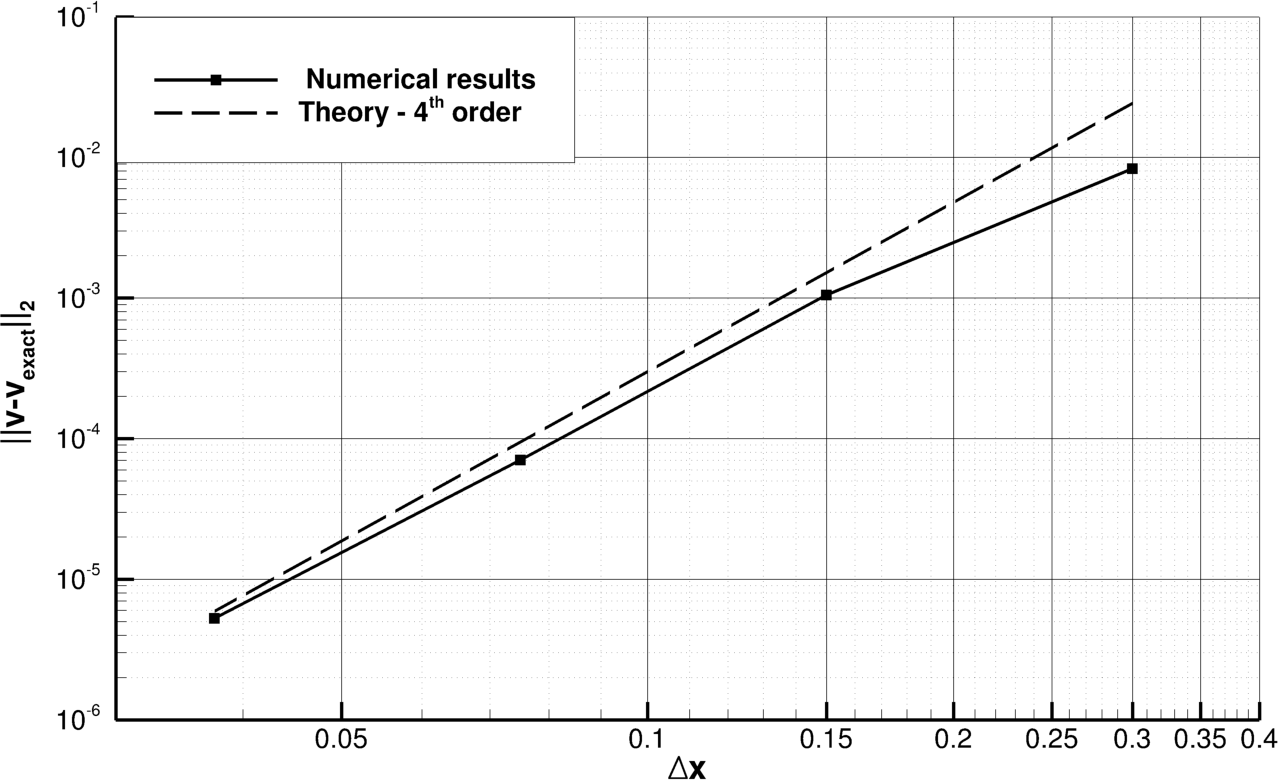}
(b)   \includegraphics[width=6.00cm,clip]{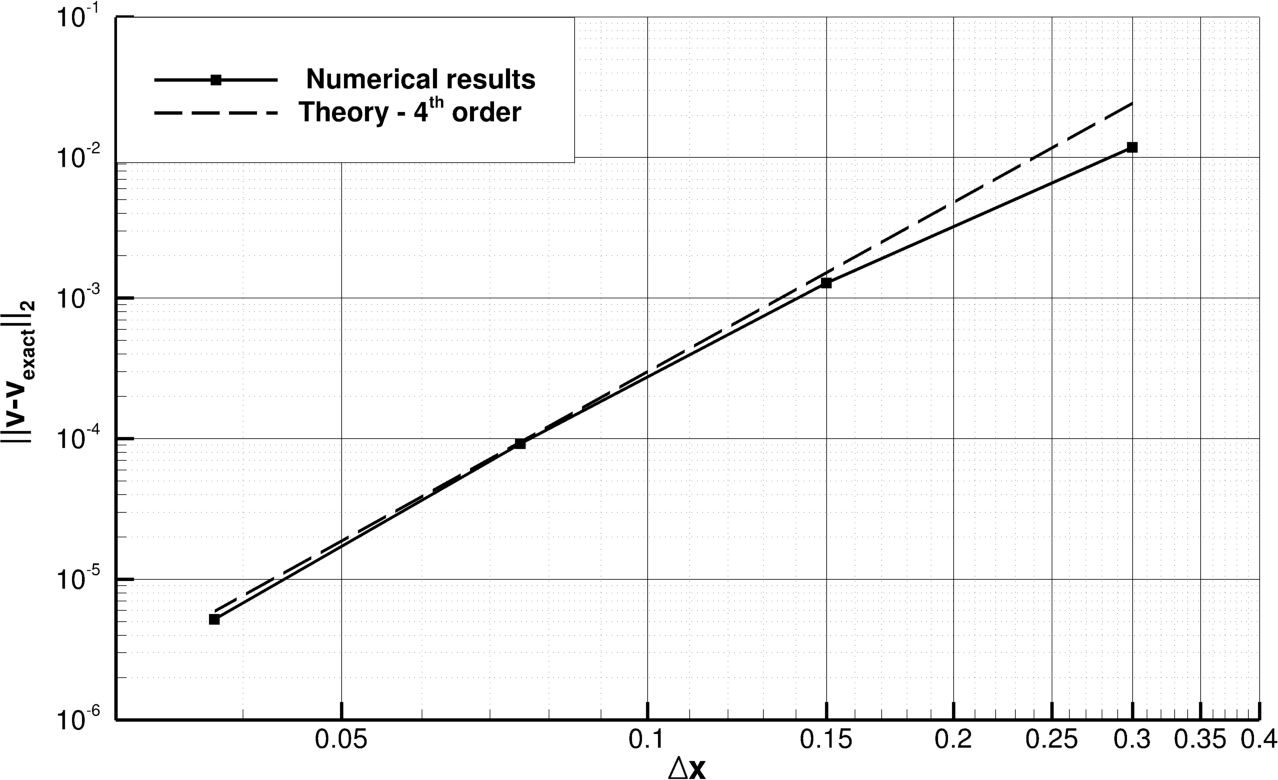}
\caption{Homentropic vortex advection: 
	$L_2$ error norm on vertical velocity as a function of grid resolution,
	for the Cartesian overlapping mesh (a) and the curvilinear overlapping mesh (b).}
\label{fig:vortex_error}
\end{figure}

The two meshes are shown in Figure~\ref{fig:vortex_mesh}.
In the illustration, the background mesh consists of \(80 \times 80\) grid points,
while the internal mesh contains \(40 \times 40\) grid points.
For Cartesian meshes (panel (a)), the internal mesh is rotated
by \(30^{\circ}\) relative to the background.
For curvilinear meshes (panel (b)), both the external and internal
meshes are smoothly distorted.
In both cases, periodic boundary conditions are applied to the
background mesh, which provides boundary conditions for the
internal meshes through the Chimera interpolation described earlier.
The numerical simulations reported here employ fourth-order accurate
central discretization and fourth-order Chimera interpolation
across non-conforming mesh boundaries. 
We used the classical fourth-order Runge-Kutta scheme for time integration.

Figure~\ref{fig:vortex_cont} presents the computed vertical
velocity contours after one flow-through time, comparing them
with the exact solution (solid black lines). On the coarsest
Cartesian mesh, with a resolution of \(L/\Delta = 40\), the vortex
exhibits significant deformation. This distortion is even more
pronounced on the curvilinear mesh at the same resolution, as expected.
However, deviations from the exact solution are negligible on the
next finer mesh, with a resolution of \(L/\Delta = 80\), for both
Cartesian and curvilinear grids.

A more detailed view is provided in Figure~\ref{fig:vortex_prof},
which shows velocity profiles across the vortex. The results confirm
that the coarsest mesh, which resolves the vortex radius with only
about six points, is inadequate for accurately representing the
vortex structure. This inadequacy leads to underestimation of the
maximum velocity and introduces oscillatory behavior in the velocity
profile. In contrast, finer grids exhibit significantly smaller
deviations from the analytical solution, demonstrating clear grid
convergence with mesh refinement.

Figure~\ref{fig:vortex_error} also quantifies grid convergence by
plotting the \(L_2\) error norm for the vertical velocity as a
function of grid resolution. The results confirm that the solver
retains full nominal fourth-order accuracy, even in the presence of
non-conforming boundaries, through the use of fourth-order
block-to-block interpolation. Use of curvilinear meshes in this 
case only yields marginally higher absolute error, but nevertheless
the asymptotic convergence rate is maintained.

\subsection{DNS of turbulent pipe flow}
\label{sec:pipeflow}

In this section, we analyze pressure-driven turbulent flow in a circular
pipe, a configuration first investigated using DNS by \citet{eggels_94}.
This flow is considered one of the most canonical cases of wall-bounded
turbulence, with extensive reference data from both experimental and
numerical studies. The computational setup adopted here is identical to
that of \citet{pirozzoli_21}. Specifically, the simulations are performed
by imposing a constant axial forcing term to mimic the effect
of a driving pressure gradient. Periodic boundary conditions are applied
along the axial direction, assuming fully developed flow, with the pipe
length set to fifteen times its radius.

Unlike the previous test case, the present computation uses an incompressible
flow model with viscosity. The viscous terms are approximated with a
second-order centered scheme, and the time derivatives are discretized using a
standard three-point backward, fully implicit formula. As a result, the
global accuracy of the scheme is second-order. Since the resulting system of
algebraic equations is fully implicit, it is solved using a multi-grid
technique and dual time stepping~\citep[see, e.g.,][]{di2014wake,
muscari2017analysis, magionesi2018modal, di2022vortex}.
To validate the numerical methodology, we simulate a flow at a low Reynolds
number (\( \Rey_b = 5300 \)), consistent with the study by \citet{eggels_94}.
Our results are compared with DNS data obtained in a cylindrical coordinate
system~\citep{pirozzoli_21}, which serve as a reference.

\begin{figure}[h]
\centering
(a) \includegraphics[width=6.00cm,clip]{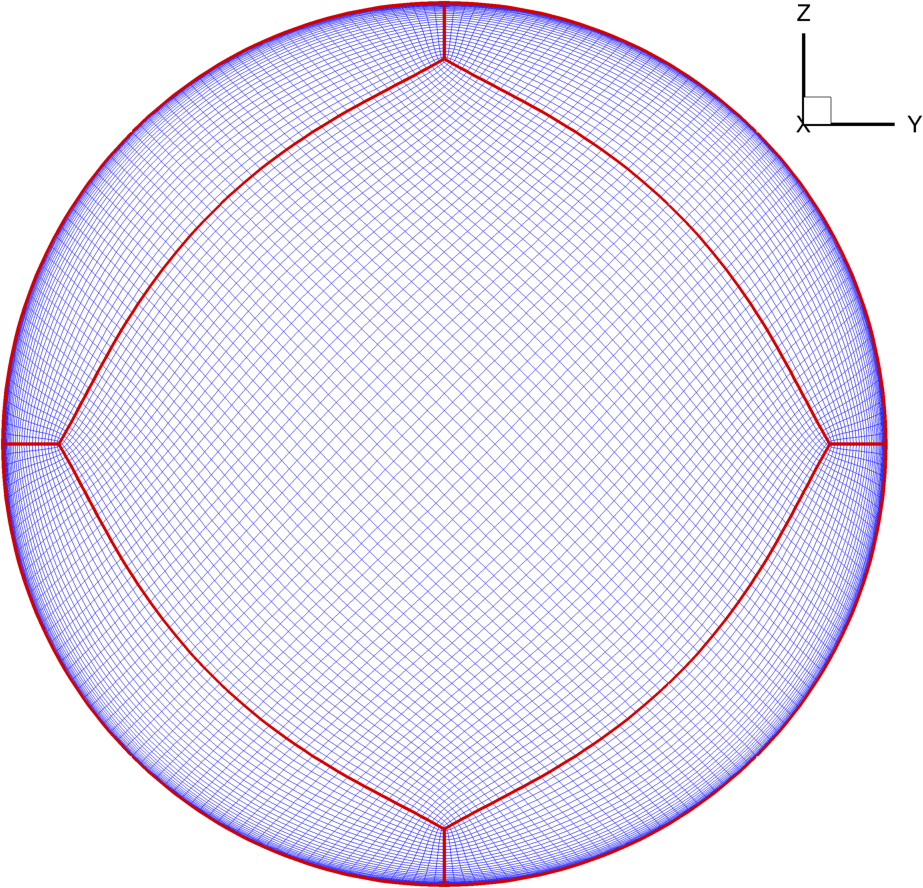}
(b) \includegraphics[width=6.00cm,clip]{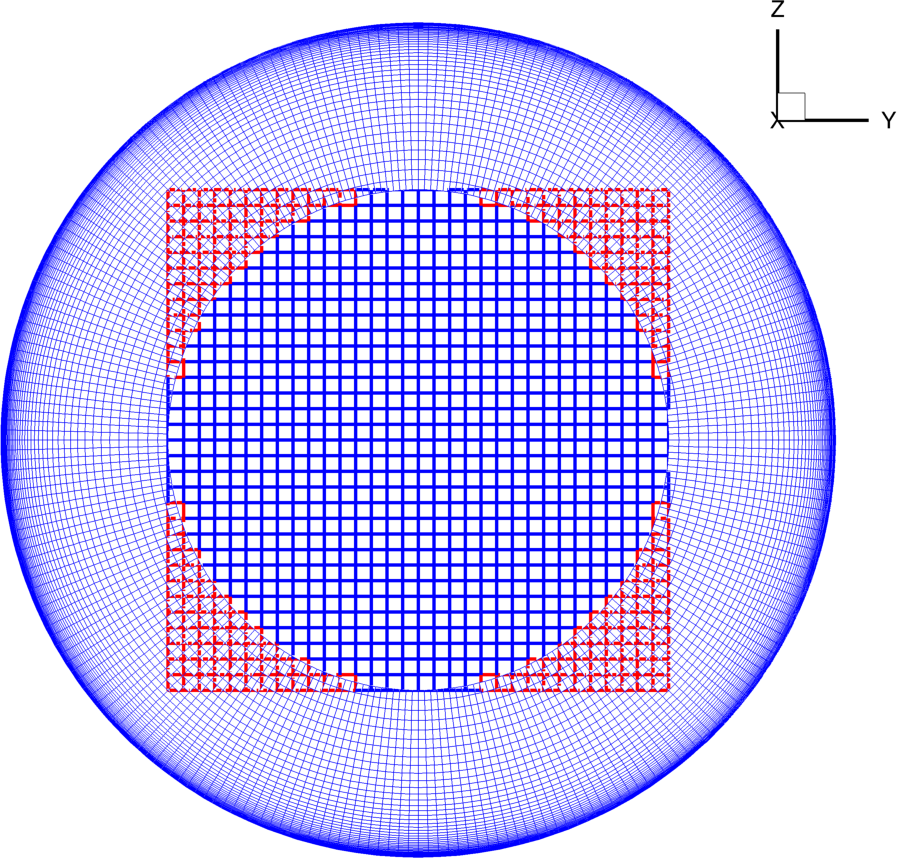}
\caption{Computational meshes for DNS of turbulent pipe flow: 
(a) block-structured mesh (edges in red); and (b) overlapping (Chimera) mesh 
(blanked cells in red).}
\label{fig:mesh}
\end{figure}

%\begin{figure}[h]
%\centering
%\includegraphics[width=10.00cm,clip]{FIGURES/CHIMERA/chimera_cf_vs_time} 
%\caption{DNS of turbulent pipe flow: time evolution of average skin friction 
%for the Chimera mesh.
%\label{fig:cf_chimera}}
%\end{figure}

This study employed two mesh configurations, as illustrated in
figure~\ref{fig:mesh}. The first configuration consists of a block-structured
mesh comprising one inner and four outer blocks arranged in an abutting
setup without overlap. Each outer block contains \(64 \times 32 \times 256\)
points in the peripheral, radial, and streamwise directions, respectively,
while the inner block contains \(65 \times 65 \times 256\) points, resulting
in a total of 3.15 million points.
The second configuration utilizes a curvilinear polar mesh for the outer
region of the pipe, comprising \(256 \times 48 \times 256\) points in the
peripheral, radial, and streamwise directions, respectively. A Cartesian
mesh is used for the pipe core, consisting of \(33 \times 33 \times 256\)
points, bringing the total to 3.4 million points. The Chimera interpolation
algorithm is employed to manage the overlap between the polar and Cartesian
regions.
For both mesh configurations, the grid points are clustered toward the solid
boundary, ensuring that the wall-normal thickness satisfies \(d^+ = d u_\tau /
\nu < 1\), where \(u_\tau\) is the friction velocity. The non-dimensional
time step is set to \(1 \times 10^{-3}\), guaranteeing a CFL number not
exceeding 0.5.
The numerical simulations were advanced in time until the pressure gradient achieved
statistical convergence. Afterwards, they were continued for approximately
one hundred flow-through times to ensure robust statistical sampling. 
%Figure~\ref{fig:cf_chimera} shows the time evolution of the ratio of the average skin friction on the pipe walls, \( F_x / (2 \pi R_{\text{pipe}} L_{\text{pipe}}) \), to \( \tau_{\text{wall}} = |dP/dx| / 2 \), which would equal 1 for steady flows. Time is made dimensionless using the bulk velocity, \( U_{\text{bulk}} \), and the pipe length, \( L_{\text{pipe}} \). Throughout the simulation, the skin friction ratio oscillates around 1, with an average value of 1.0016.

\begin{figure}
\centering
(a) \includegraphics[width=6.00cm,clip]{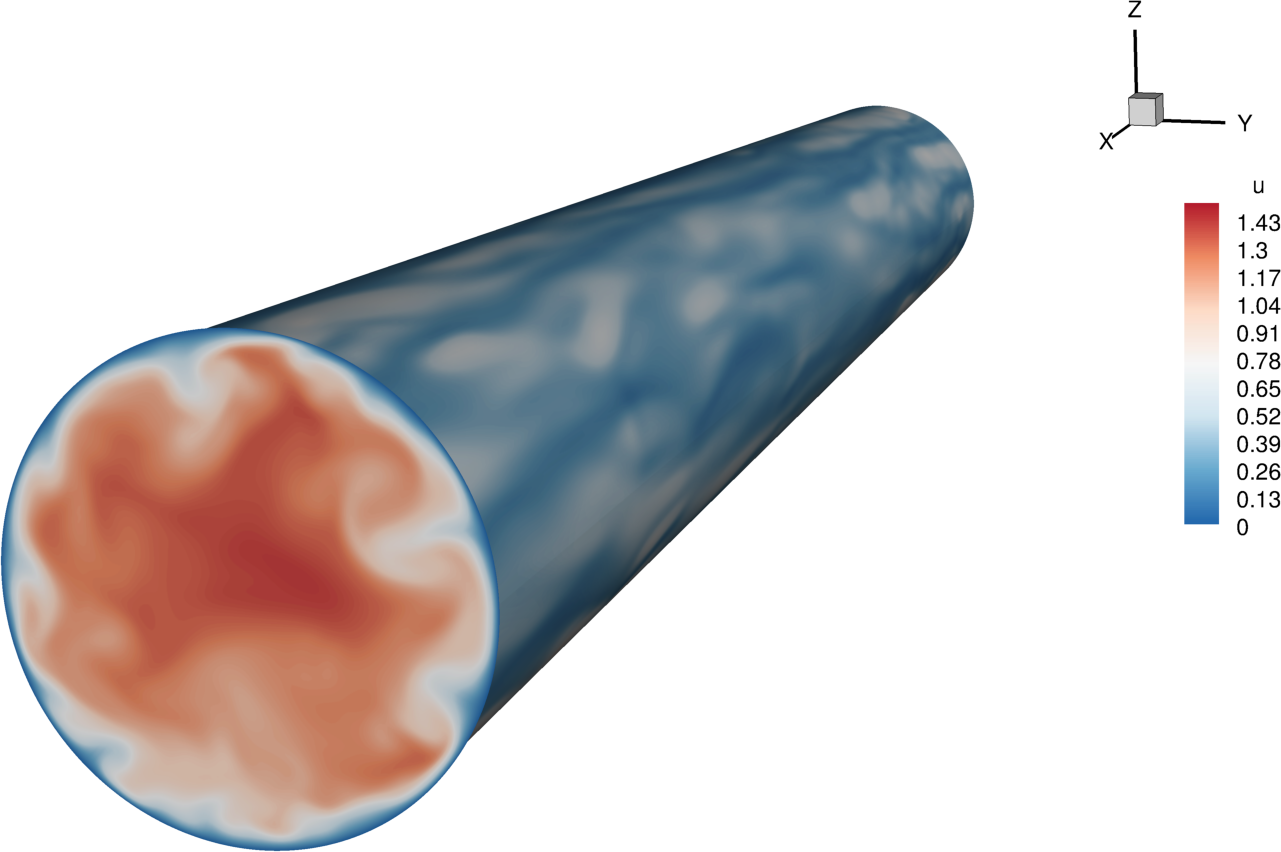}
(b) \includegraphics[width=6.00cm,clip]{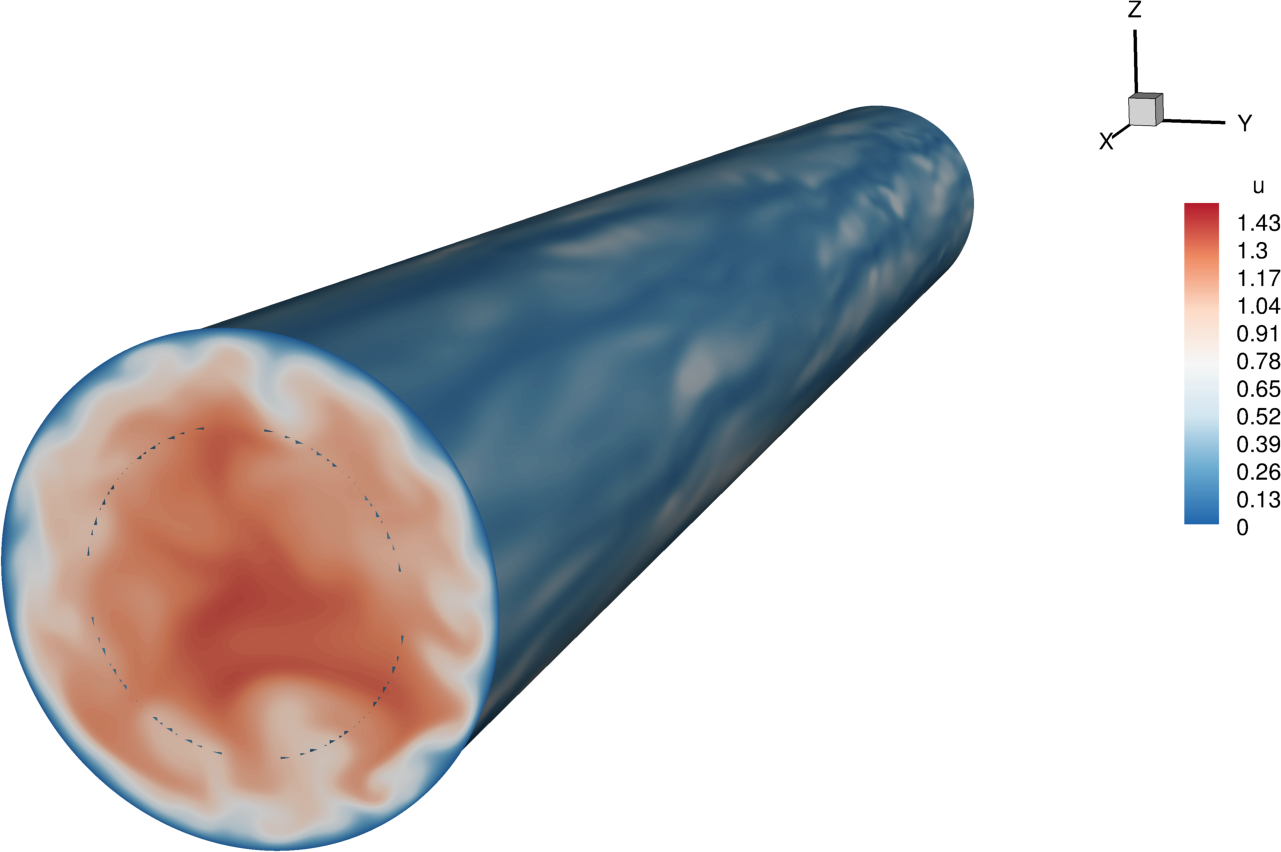}
\caption{DNS of turbulent pipe flow: instantaneous streamwise velocity fields obtained
with a block-structured mesh (a) and a Chimera mesh (b).
\label{fig:u}}
\end{figure}

Figure~\ref{fig:u} shows instantaneous axial velocity contours  
in both a cross-stream plane and a near-wall plane. The flow exhibits  
the characteristic features of pipe turbulence, including near-wall  
streaks with alternating regions of high- and low-speed momentum  
(depicted in red and blue, respectively). The cross-stream plane  
highlights large-scale sweeps and ejections, which play a critical  
role in momentum exchange.  
Notably, while the velocity fields are nominally sampled at the same  
time instants, discrepancies arise due to the sensitive dependence  
of instantaneous turbulent flow realizations on initial conditions  
and mesh geometry. 

\begin{figure}
\centering
\includegraphics[width=6cm,clip]{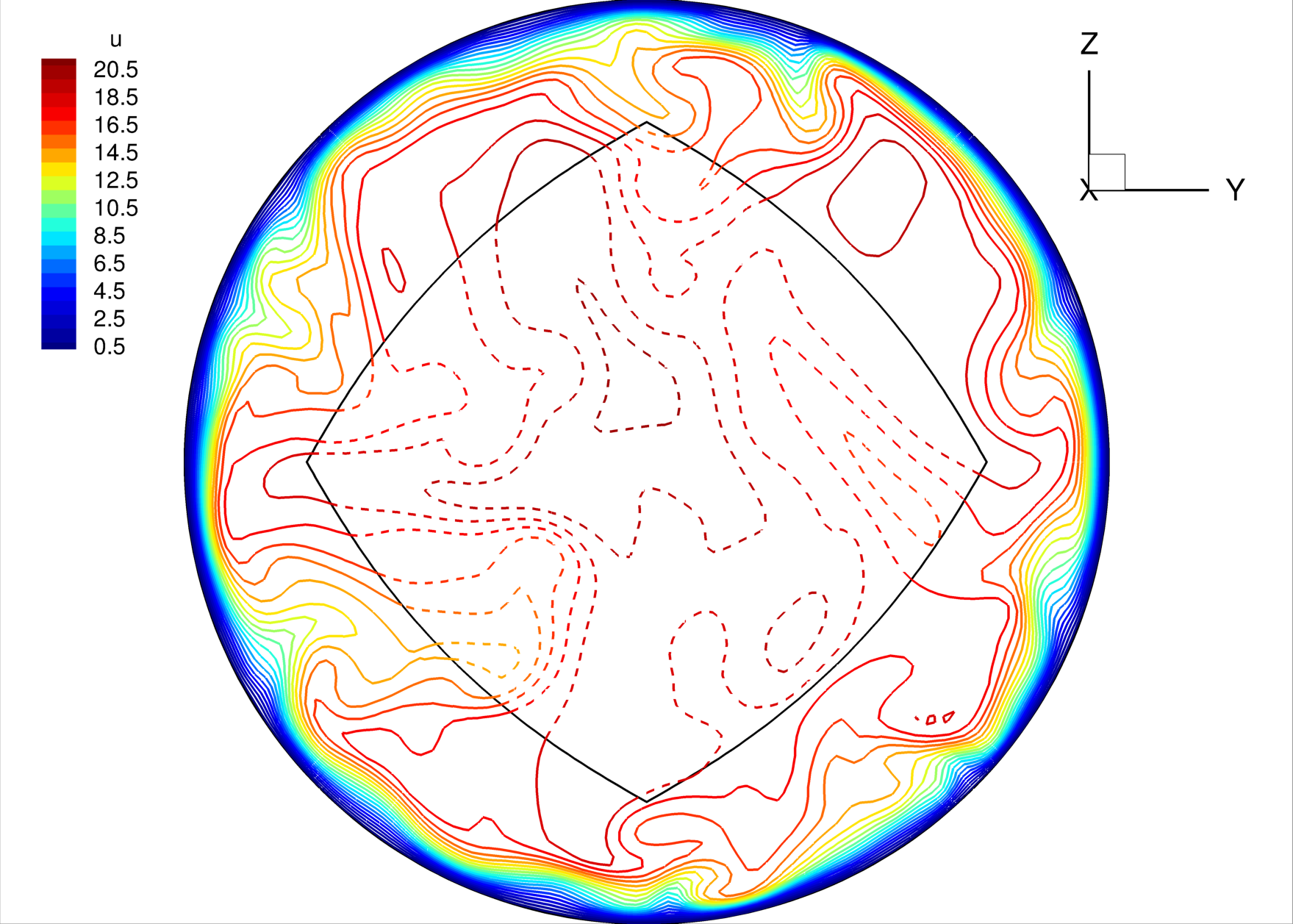}
\includegraphics[width=6cm,clip]{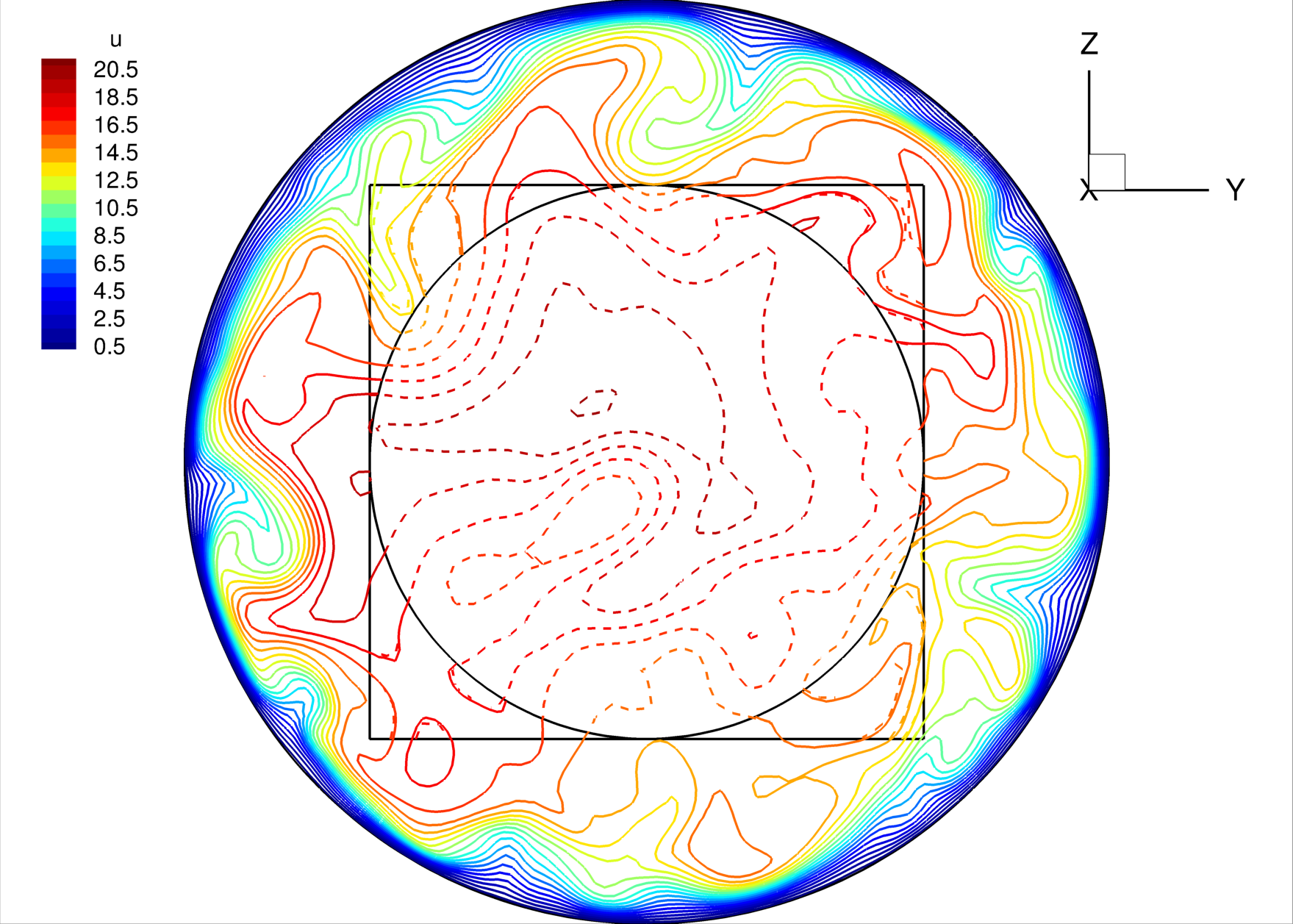}
\includegraphics[width=12cm,clip]{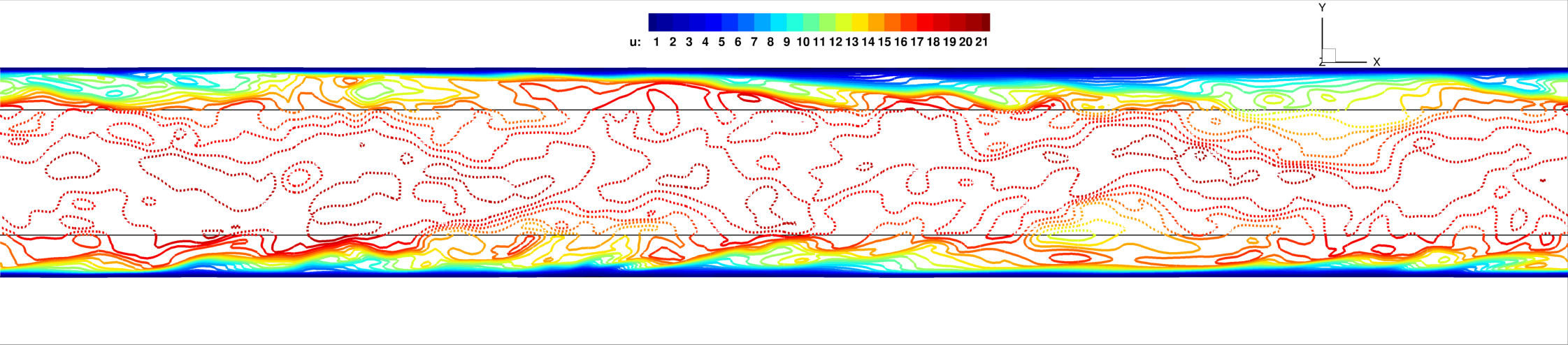}
\caption{DNS of turbulent pipe flow: instantaneous streamwise velocity fields 
in a cross-stream plane obtained with a block-structured mesh (a) and a Chimera mesh (b);
and longitudinal slice of instantaneous velocity field obtained with a Chimera mesh (c).
Solid lines indicate velocity contours in the outer mesh, and
dashed lines velocity contours in the inner mesh.
\label{fig:uzy}}
\end{figure}

Figure~\ref{fig:uzy} displays iso-lines of the streamwise velocity. The 
cross-stream slices reveal that the axial velocity is perfectly smooth in 
the case of the abutting mesh (panel a). Similarly, excellent continuity is 
achieved with the Chimera mesh (panel b). Notably, in the overlapping 
region between the outer mesh (solid lines) and the inner mesh (dashed 
lines), the velocity contours align closely, demonstrating the effectiveness 
and accuracy of the interpolation procedure.
Further validation is provided in panel (c), which shows a longitudinal 
slice of the pipe for the Chimera mesh. As with the previous results, no 
visible discontinuities or numerical artifacts, such as wiggles, are observed 
across the transition region between the inner and outer meshes. 

\begin{figure}
\centering
(a)   \includegraphics[width=6.00cm,clip]{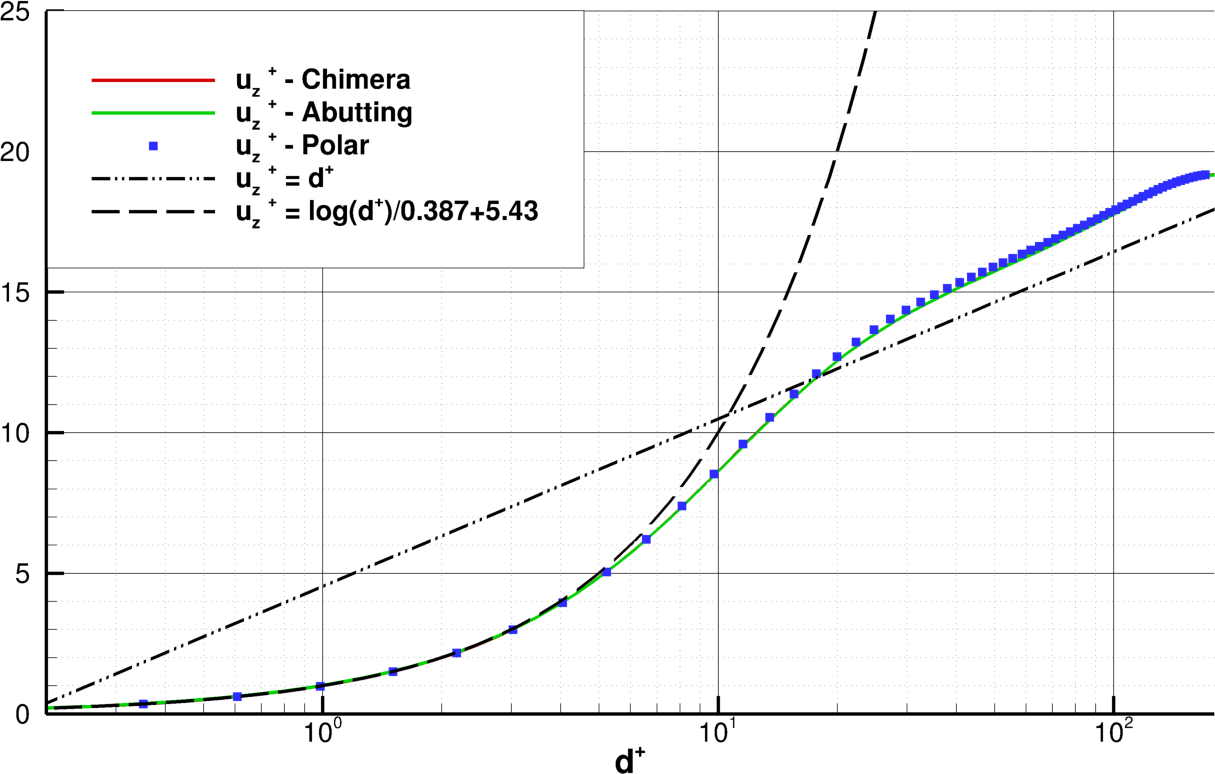}
(b)   \includegraphics[width=6.00cm,clip]{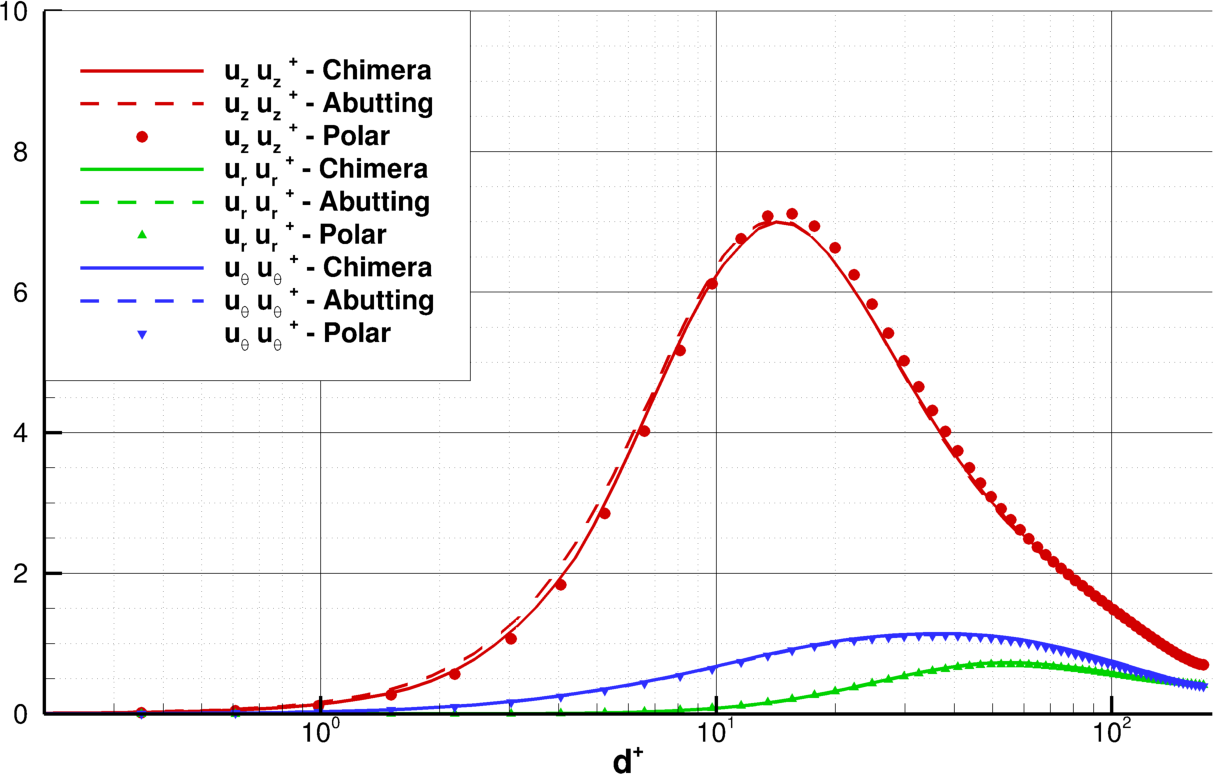}
\caption{DNS of turbulent pipe flow: comparison of mean velocity profiles (a), and velocity
variances, for the case of cylindrical coordinates~\citep[symbols][]{pirozzoli_24},
abutting mesh (green) and Chimera mesh (red).
\label{fig:u_comp}}
\end{figure}

Figure~\ref{fig:u_comp} provides a quantitative evaluation of the flow statistics,
presenting the mean velocity and velocity variance profiles obtained from DNS on both
abutting and Chimera meshes. These results are compared with DNS data computed in cylindrical
coordinates \citep{pirozzoli_24}. The profiles show excellent agreement across all statistics,
with no signs of localized spurious increases or reductions in the intensity of turbulent fluctuations. 

\begin{figure}
\centering
\begin{tabular}{cc}
(a) & \includegraphics[width=11.00cm,clip]{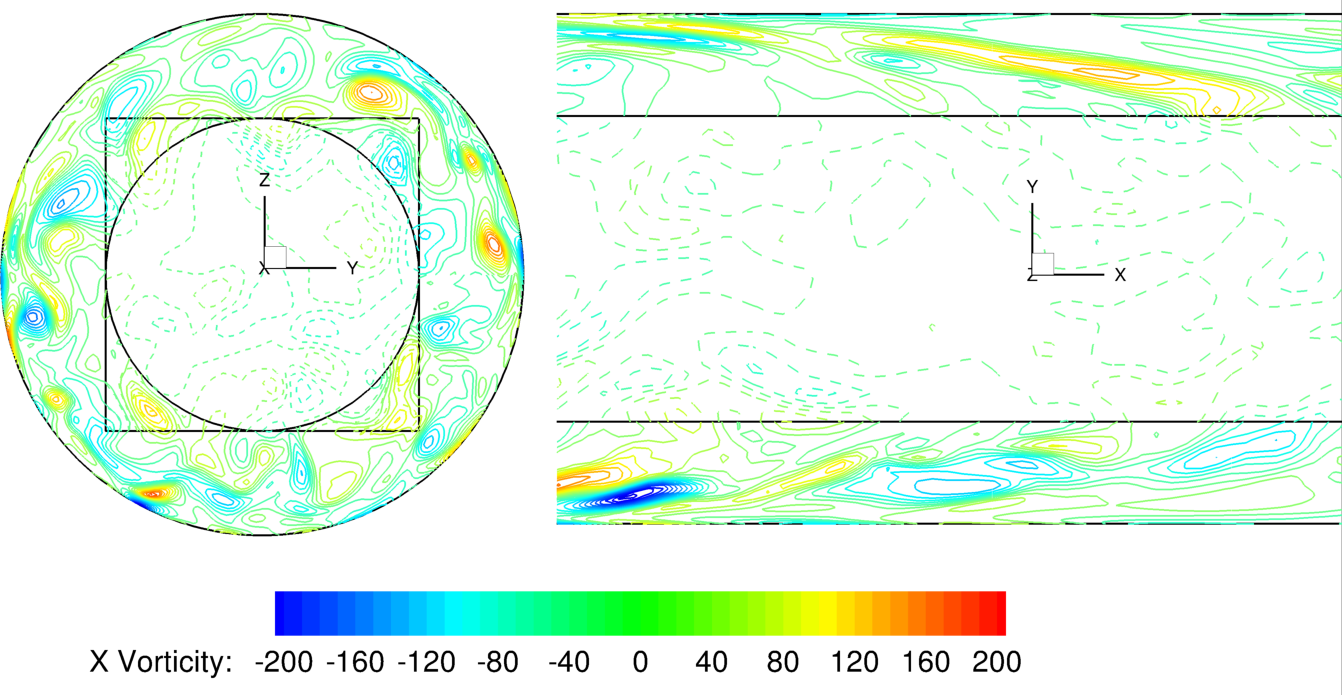}
\\*[0.5cm]
(b) & \includegraphics[width=11.00cm,clip]{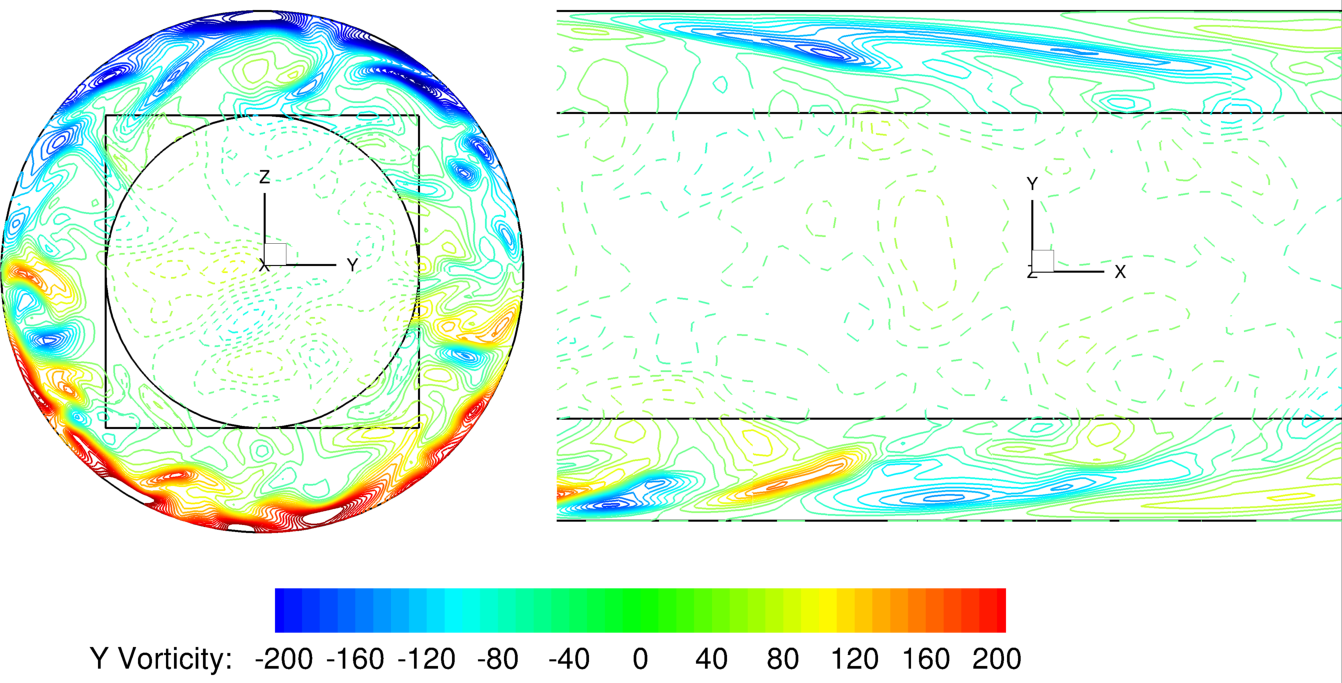}
\\*[0.5cm]
(c) & \includegraphics[width=11.00cm,clip]{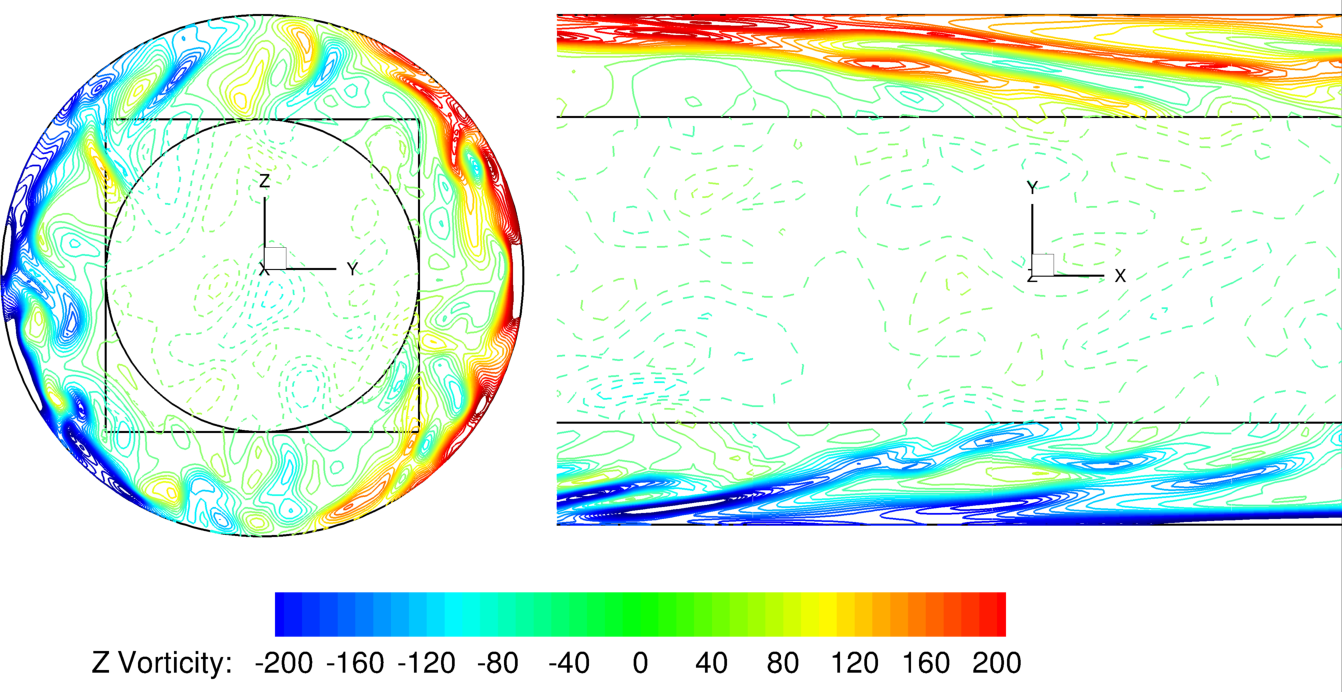}
\end{tabular}
\caption{DNS of turbulent pipe flow on Chimera mesh:
instantaneous vorticity components 
in cross-stream (left) and longitudinal (right) planes. Top to bottom:
x-component (a), y-component (b), z-component (c).
\label{fig:vorticity}}
\end{figure}

As an additional verification, figure~\ref{fig:vorticity} presents iso-lines of the
vorticity components, providing insight into potential errors in capturing small-scale turbulence
across meshes with different resolutions. Vorticity, which involves derivatives of the velocity components,
is a key metric for assessing the strength of turbulence at small scales~\citep{moin_95}.
Once again, no evidence of discontinuities or numerical artifacts, such as jumps or wiggles,
is observed near the mesh interfaces. Notably, the strong shear layers originating from the pipe wall
are accurately transmitted into the interior mesh. Furthermore, the vorticity contours in the overlap region
between the two meshes (depicted with solid and dashed lines) show excellent conformity,
demonstrating the robustness and accuracy of the interpolation algorithm. 

\subsection{High-fidelity simulation of ship propeller wake}
\label{sec:propeller}

\begin{figure}
\begin{tabular}{cc}
\centering
\includegraphics[width=6.00cm,clip]{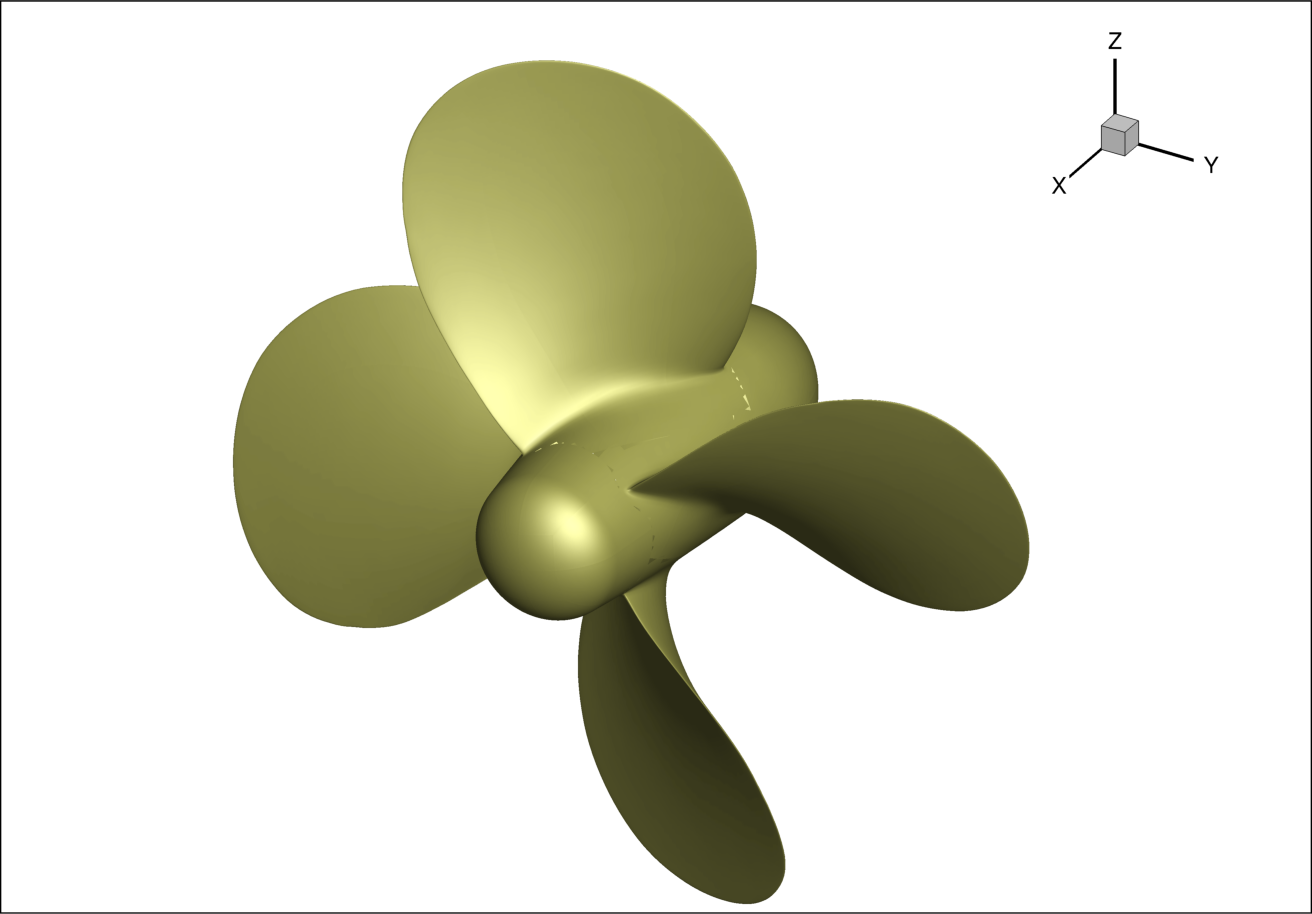}
&
\includegraphics[width=6.00cm,clip]{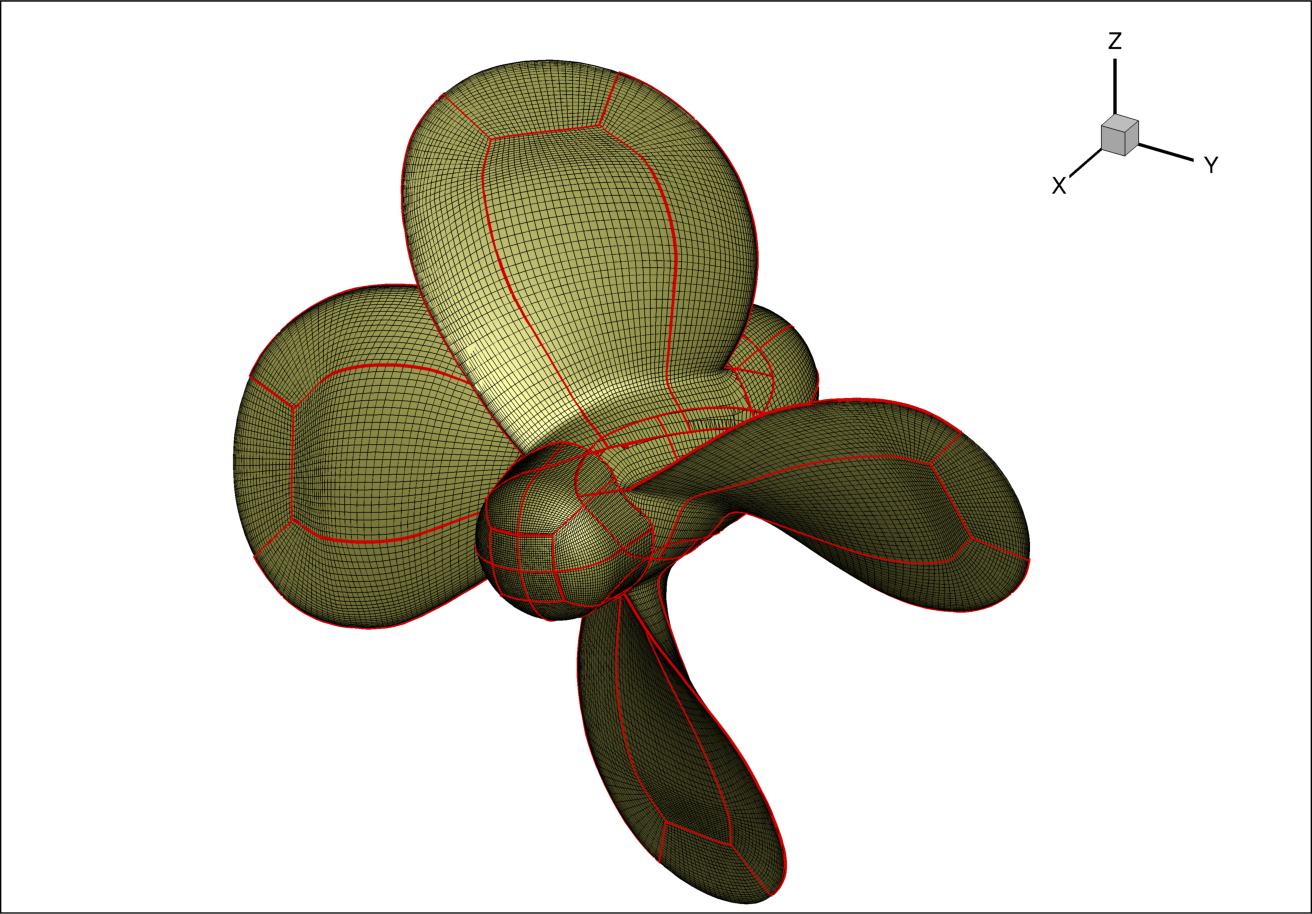}
\\
(a) & (b)
\\
\includegraphics[width=6.00cm,clip]{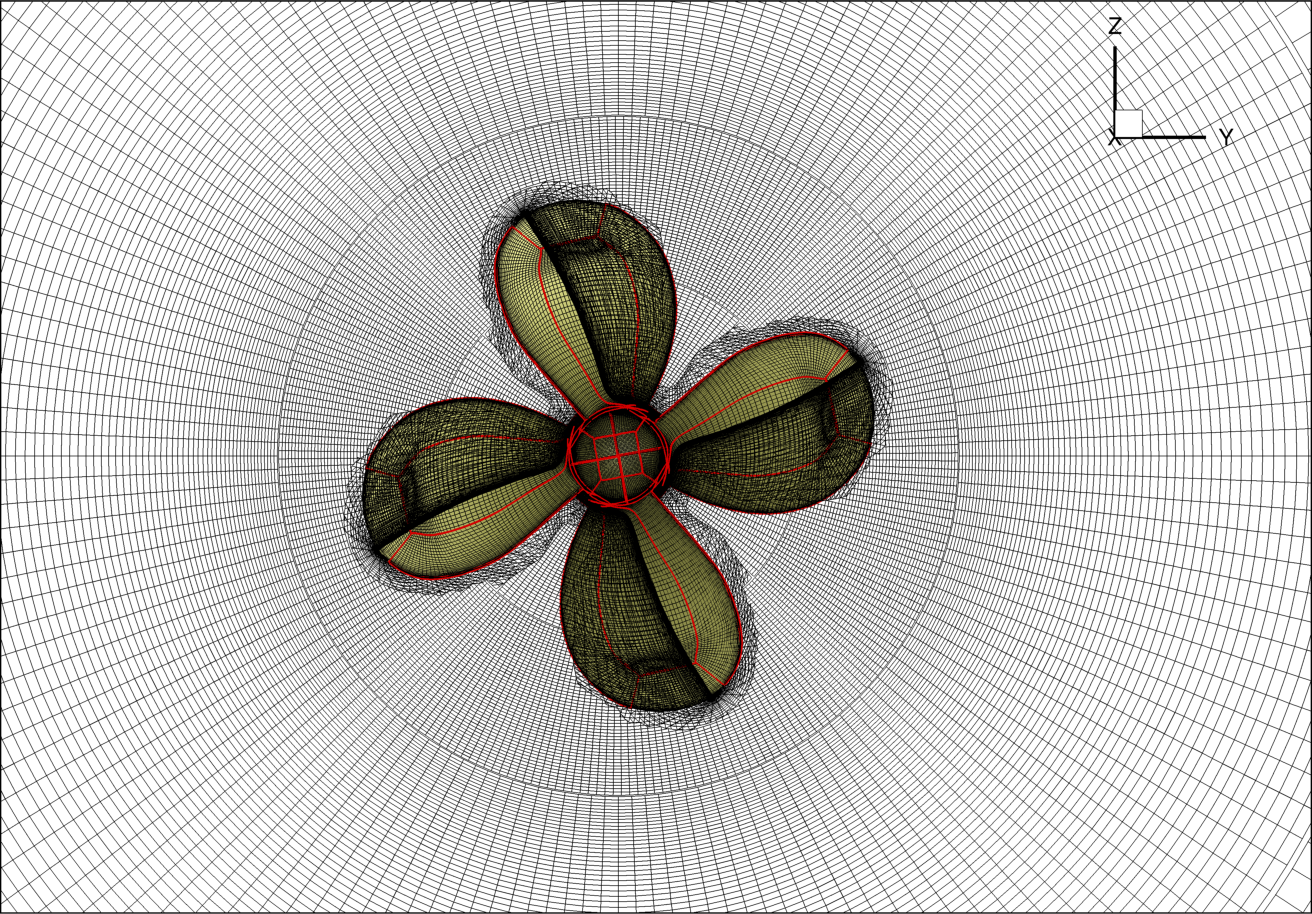}
&
\includegraphics[width=6.00cm,clip]{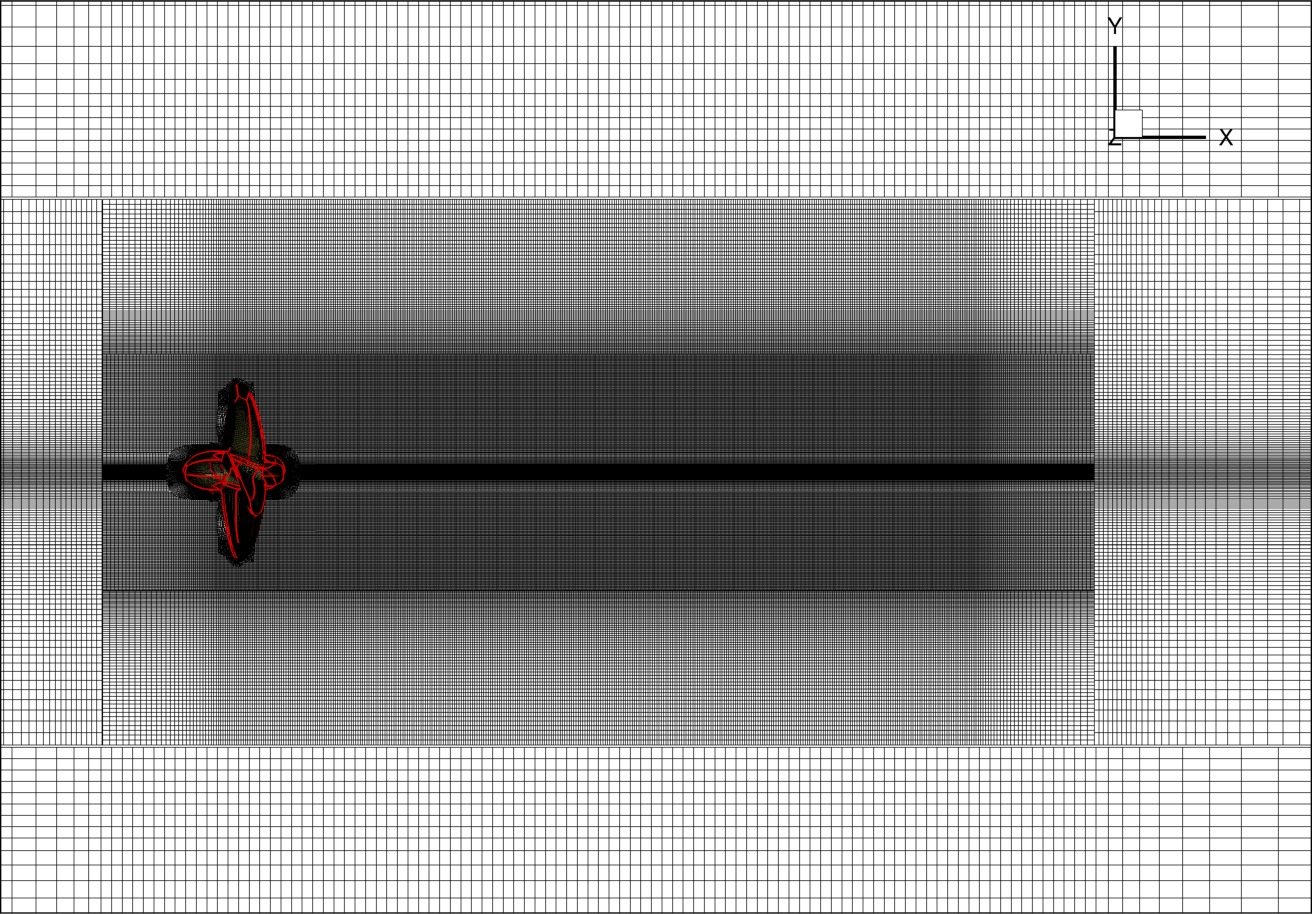}
\\
(c) & (d)
\end{tabular}
\caption{INSEAN propeller E779A: (a) model rendering; (b) multi-block surface grid; (c) mesh cross-section on plane $x=0$; (d) mesh cross--section on plane $z=0$.}
\label{fig:prop_model}
\end{figure}

As a final example of application to actual engineering problems, we performed
the computation of the flow past an isolated rotating propeller in mild loading
conditions, for which both experimental data~\cite{felli2011mechanisms} and
numerical simulation are available~\cite{muscari2013modeling}. 
Figure~\ref{fig:prop_model} reports the geometry of the stock propeller INSEAN
E779A, the surface grid, and two sections of the volume grid on the planes
\( x=0 \) and \( z=0 \). The complete grid consists of 501 partially overlapping
blocks, for a total of approximately 21 million points. 

We performed the computation in the rotating reference frame as a Detached
Eddy Simulation (DES)~\cite{spalart1997comments} for a loading condition given
by \( J = U_\infty / (n D) = 0.71 \), where \( J \) is the advance ratio, \( U_\infty \)
is the speed of the undisturbed flow, \( n \) is the number of propeller revolutions
per second, and \( D \) is the propeller diameter.

\begin{figure}
\centering
\begin{tabular}{cc}
%%%%%%%%%%%%%%%%%%%%%%%%%%%%%%%%%%%%%%%%%%%%%
\setlength{\tabcolsep}{10pt} % Default value: 6pt
\renewcommand{\arraystretch}{1.5} % Default value: 1
\begin{tabular}[b]{c|c|c}
\hline
   &    {\small $K_T$}   &  {\small $K_Q$}  \\
\hline \hline 
{\small Med.} & $0.223$  & $0.0471$ \\
{\small Fine} & $0.226$    & $0.0420$  \\
\hline 
{\small Exp.}  & $0.221$  & $0.0425$
\\
\hline 
\end{tabular}
%%%%%%%%%%%%%%%%%%%%%%%%%%%%%%%%%%%%%%%%%%%%%
&
\includegraphics[height=4.00cm,clip]{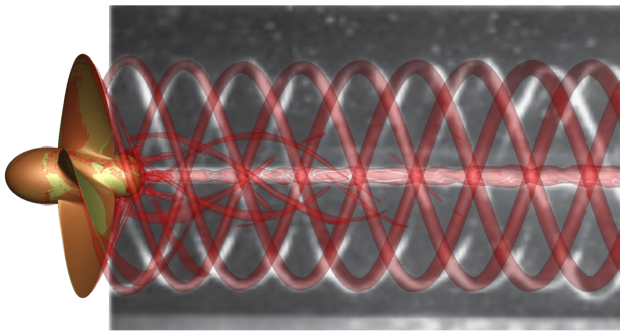}
\end{tabular}
\caption{INSEAN propeller E779A; $J=0.71$: 
Left: thrust and torque coefficient computed on the fine and medium grid;
comparison with experimental data \cite{felli2011mechanisms}. Right: Computed and observed tip vortices}
\label{fig:prop_forces}
\end{figure}

To evaluate numerical uncertainty, we carried out simulations at two grid levels.
Specifically, the coarser grid was generated by removing every other grid point
from the finer grid in each direction. The table on the left of Figure~\ref{fig:prop_forces}
compares the computed thrust and torque coefficients on both grid levels with the experimental
measurements. The right part of the figure compares the computed tip vortices, shown in red
and visualized as suggested by~\cite{jeong1995identification}, with the experimental tip vortices,
visualized by inducing vortex core cavitation. The agreement between numerical and experimental
data is very satisfactory regarding both force coefficients and wake topology. 

\begin{figure}
\begin{tabular}{cc}
\centering
(a) &
\includegraphics[width=12.00cm,clip]{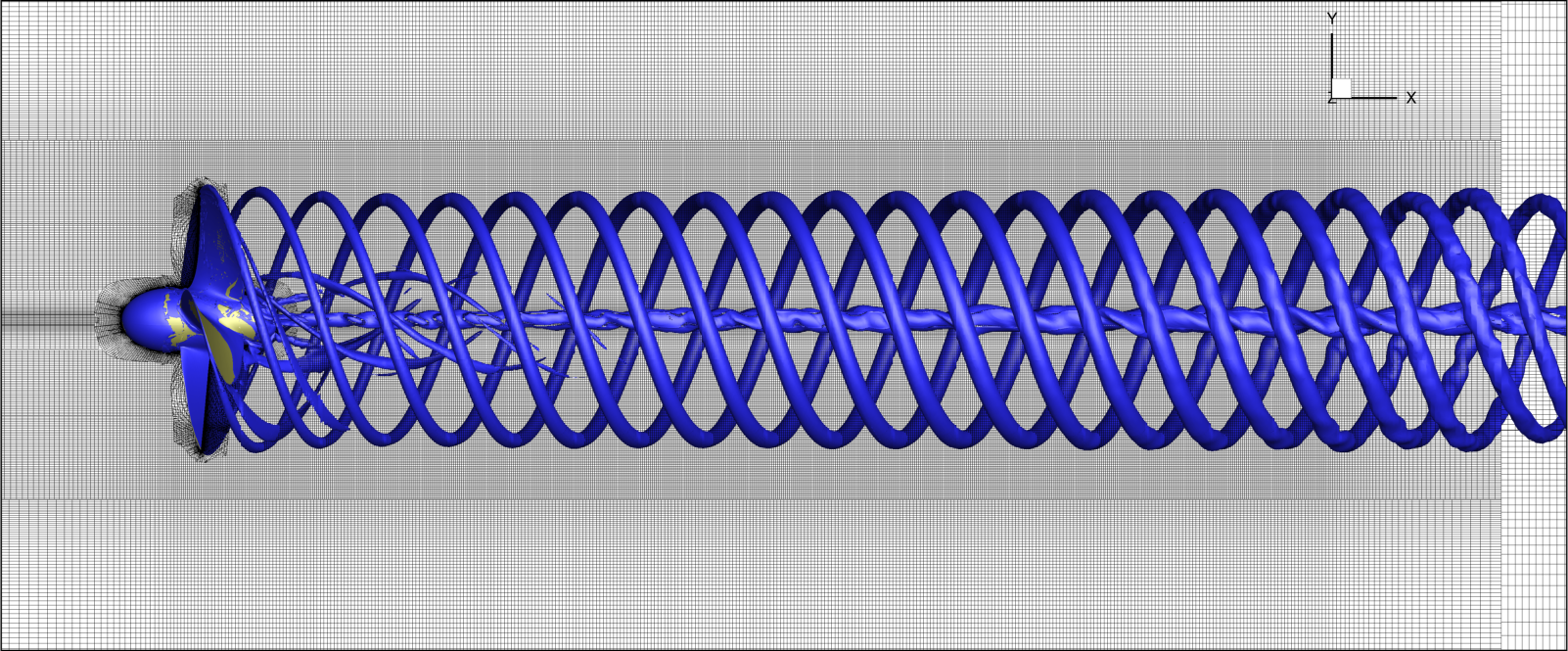}
\\
%(b) &
%\includegraphics[width=12.00cm,clip]{FIGURES/PROPELLER/Xvorticity}
%\\
(b) &
\includegraphics[width=12.00cm,clip]{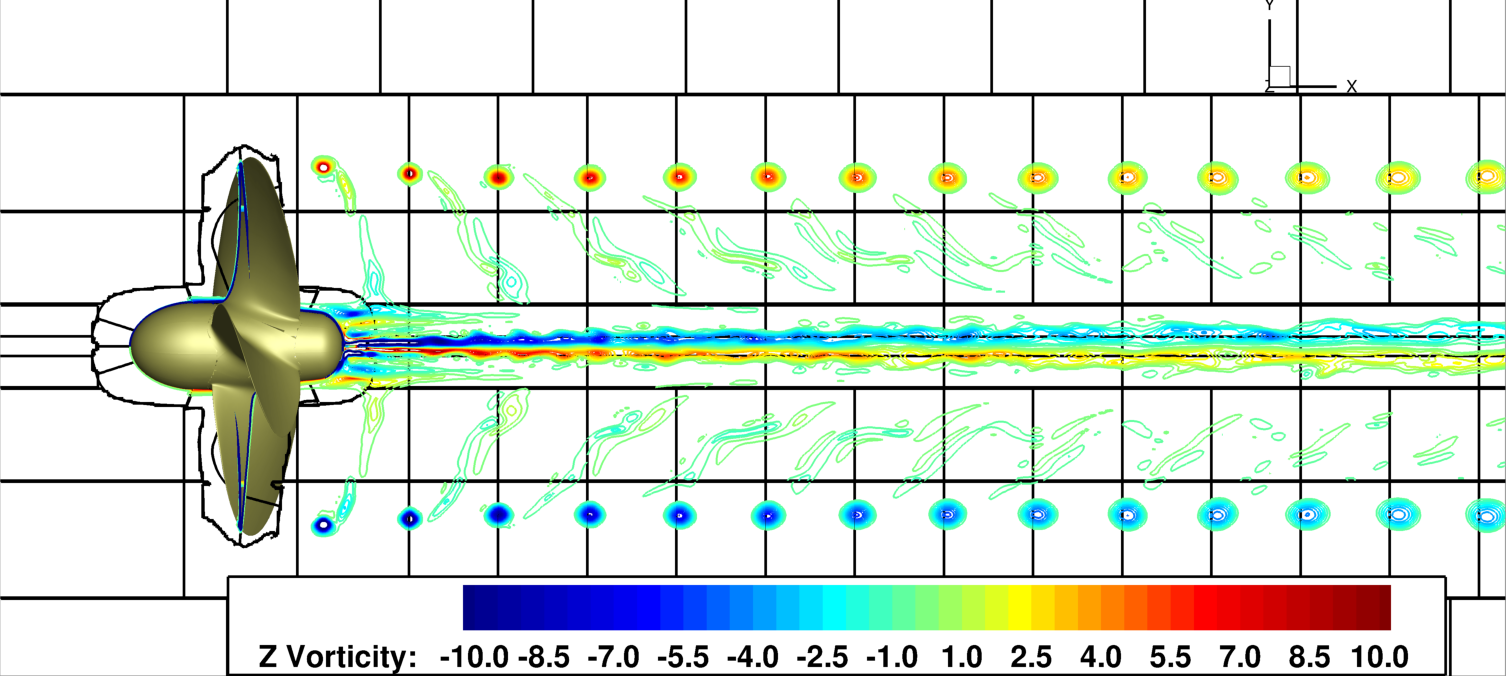}
\\
(c) &
\includegraphics[width=12.00cm,clip]{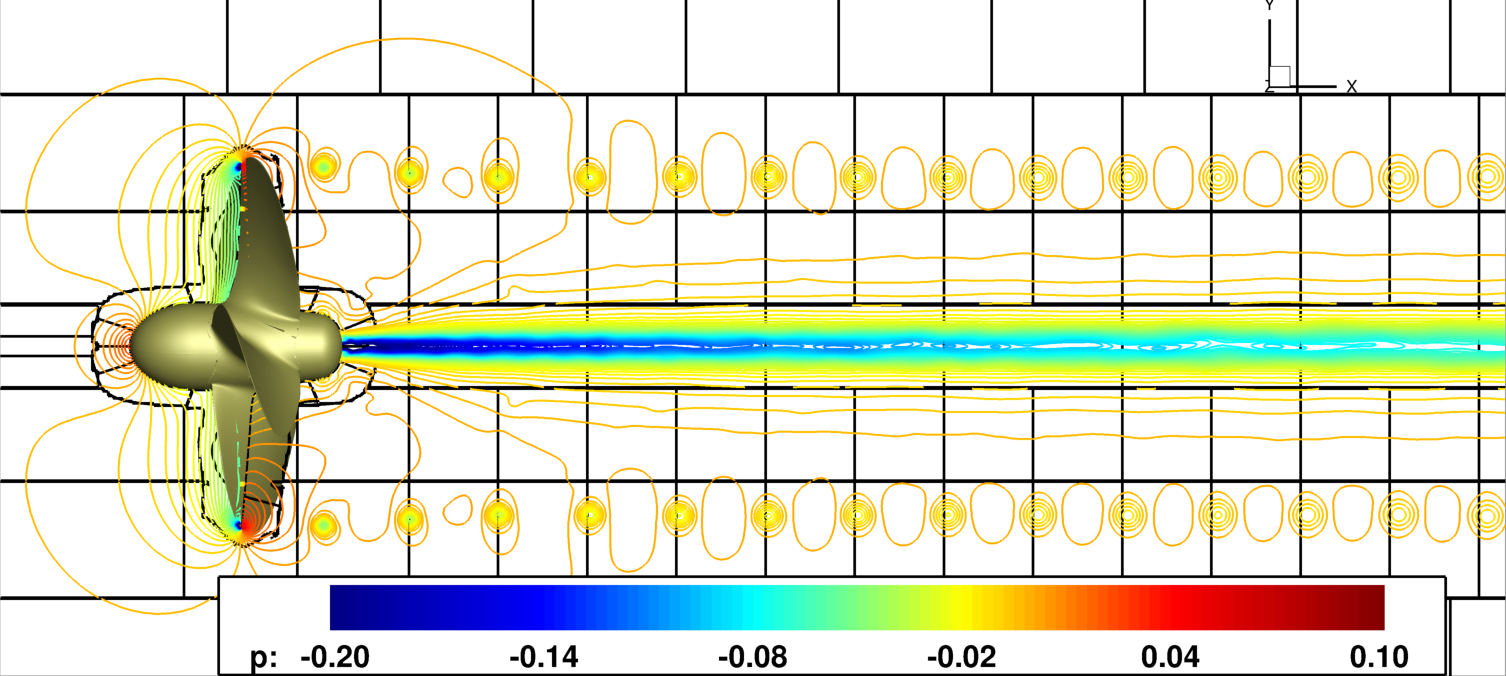}
\end{tabular}
\caption{INSEAN propeller E779A; $J=0.71$.
(a) tip vortices identified by the $\lambda_2$ criterion with $\lambda_2=-1$;
(b) Z component of vorticity on the z=0 plane;
(c) pressure field on the z=0 plane. 
The solid black lines denote the boundaries of the mesh blocks.
\label{fig:prop_vorticity}}
\end{figure}

Finally, Figure~\ref{fig:prop_vorticity} presents the solution in terms of
tip vortices, identified as suggested in~\cite{jeong1995identification}
with \( \lambda_2 = -1 \), along with the vorticity field on the plane
\( z = 0 \). The bottom part of the figure shows the axial velocity and
pressure fields.
The figure clearly demonstrates that the tip vortex can persist while
gradually thickening over a long distance downstream of the propeller,
due to the near absence of numerical dissipation. This behavior would not
be possible with conventional, non-energy-conserving algorithms.
We would like to particularly emphasize that, as with the flow in the
circular pipe, the solution develops smoothly without discontinuities
across the overlapping blocks, even in the far field where the grid is
coarsened to save CPU time. 

\section{Conclusions}
\label{sec:conclusions}

We have presented a computational framework for high-fidelity simulations of turbulent flows based on the Chimera approach. This methodology employs multiple overset meshes, potentially with varying resolutions and/or coordinate systems, to model complex geometries and flow phenomena effectively. These meshes can either move relative to one another or remain static.
The proposed computational approach utilizes a central discretization applied to the split form of the convective terms in the Navier-Stokes equations, enabling the discrete preservation of total kinetic energy even on smoothly distorted curvilinear meshes. Minimal filtering is applied for strongly distorted meshes to mitigate the emergence of dispersive errors that could compromise the numerical solution while preserving the formal order of accuracy of the underlying scheme.
A key feature of the solver is its ability to discretely preserve free-stream properties without disturbances discretely, achieved through a three-dimensional generalization of the method proposed by \citet{visbal_02}, as detailed in~\ref{sec:preservation}.
The core of the Chimera algorithm relies on a fast and efficient donor cell search mechanism across overlapping meshes, coupled with high-accuracy interpolation schemes of up to fourth order. Information exchange between meshes is seamlessly facilitated by augmenting the Navier-Stokes equations with appropriate sponge terms, preventing the onset of spurious saw-tooth oscillations and ensuring smooth transitions across mesh interfaces.

Numerical simulations of academic benchmark problems with known analytical solutions have demonstrated that the proposed Chimera algorithm retains formal accuracy up to fourth order, even with partially overlapping meshes. We also validated the algorithm for turbulent flow in a pressure-driven pipe. Numerical results reveal no spurious oscillations in large-scale properties (e.g., velocity) or small-scale properties (e.g., vorticity), with smooth transitions across non-matching blocks.
The resulting flow statistics agree with reference results obtained using standard structured solvers. To the best of our knowledge, this represents the first successful validation of DNS for turbulent flows using an overlapping mesh solver.
Finally, we have applied the method to a test case relevant to marine engineering involving a rotating propeller. We performed a Detached Eddy Simulation to analyze the wake structure for this case. The properties of the numerical algorithm enable accurate prediction of the tip vortex, with its strength preserved over a significant distance downstream of the propeller hub. The numerical results show strong qualitative and quantitative agreement with available experimental data, further validating the robustness and accuracy of the proposed approach.

The method outlined in this paper opens the door for DNS, LES, and DES of turbulent flows in complex geometries, a task that the lack of efficient computational techniques for monolithic meshes has been previously limited. A key advantage of the Chimera approach is that it allows the split of the computational domain into separate, overlapping grids designed to align optimally with specific geometric features or flow regions. This flexibility allows one to generate each mesh independently, making it easier to handle complex geometries and multi-scale problems.
Chimera meshes are especially valuable for problems involving moving boundaries, such as rotating blades, aircraft in flight, or fluid-structure interactions, where traditional single-mesh approaches would face significant challenges.

\section*{Acknowledgments}
This research received financial support from ICSC-Centro Nazionale di Ricerca in
`High Performance Computing, Big Data and Quantum Computing', funded by European Union-NextGenerationEU. The results reported in this paper have been achieved
using the EuroHPC Research Infrastructure resource LEONARDO based at CINECA,
Casalecchio di Reno, Italy.

\appendix
\section{Computation of metric tensor for uniform flow preservation} \label{sec:preservation}

It is easy to check that in the case of a uniform flow $\bar{\rho},\bar{\vecu},\bar{e}$, equation \eqref{vect_eq}  
yields
\begin{equation}
\left[ 
\begin{array}{c}
\displaystyle
\bar{\rho} \bar{\vecu} \cdot \sum_{j=1}^d \sum_{\ell=1}^L a_\ell 
                    { \left[ (J \nabla \xi_j)_{\N;\ell} 
                           - (J  \nabla \xi_j)_{\N;-\ell} \right]_\N }   
\\*[3mm]
\displaystyle
\bar{\rho} \bar{\vecu} \bar{\vecu} \cdot \sum_{j=1}^d \sum_{\ell=1}^L a_\ell 
                    { \left[ (J \nabla \xi_j)_{\N;\ell} 
                           - (J  \nabla \xi_j)_{\N;-\ell} \right]_\N }   
+ \bar{p} 
\sum_{j=1}^d \sum_{\ell=1}^L a_\ell
                    { \left[ (J \nabla \xi_j)_{\N;\ell} 
                           - (J  \nabla \xi_j)_{\N;-\ell} \right]_\N }   
\\*[3mm]
\displaystyle
\bar{\rho} \bar{H} \bar{\vecu} \cdot \sum_{j=1}^d \sum_{\ell=1}^L a_\ell 
                    { \left[ (J \nabla \xi_j)_{\N;\ell} 
                           - (J  \nabla \xi_j)_{\N;-\ell} \right]_\N }   
\end{array}
\right] = 0 .
\label{uniform}
\end{equation}
Therefore, uniform flow preservation requires that
\begin{equation}
\sum_{j=1}^d \sum_{\ell=1}^L a_\ell
                    \left[  (J \nabla \xi_j)_{\N;\ell} 
                     - (J  \nabla \xi_j)_{\N;-\ell}  \right]    = 0,
\label{uniform_tens}
\end{equation}
i.e., in three dimensions at $(i,j,k)$,
\begin{equation}
\begin{array}{llll}
\sum_{l=1}^L a_l &
\displaystyle
               \left[  (J \nabla \xi)_{i+l,j,k} 
                     - (J \nabla \xi)_{i-l,j,k}  \right] & + &
\\*[3mm]
\sum_{m=1}^L a_m &
\displaystyle
               \left[  (J \nabla \eta)_{i,j+m,k} 
                     - (J \nabla \eta)_{i,j-m,k}  \right]    & + & 
\\*[3mm]
\sum_{n=1}^L a_n &
\displaystyle
               \left[  (J \nabla \zeta)_{i,j,k+n} 
                     - (J \nabla \zeta)_{i,j,k-n}  \right]    & = & 0 ,
\end{array}
\label{discr_uf}
\end{equation}
which is a finite-difference approximation of the identity $\partial (J\xi_{s,r}) / \partial \xi_s = 0$, with $r=1,2,3$.

We can discretize this expression with the finite-difference approximation \eqref{fin_diff} of the 
exact expressions 
\begin{equation}
\displaystyle
J \nabla \xi_p = \frac{\varepsilon_{pqr}}{2}
 \frac{\partial \vecx}{\partial \xi_q} \times \frac{\partial \vecx}{\partial \xi_r} ,
\label{exact_tensor}
\end{equation}
$\varepsilon_{pqr}$ being the Levi-Civita tensor. For instance, at $\vecx_{i,j,k}$ one would have
\begin{equation}
\left. J \nabla \xi \right|_{i,j,k} = \sum_{m=1}^L a_m \frac{(\vecx_{i,j+m,k}-\vecx_{i,j-m,k})}{h} \times
                                      \sum_{n=1}^L a_n \frac{(\vecx_{i,j,k+n}-\vecx_{i,j,k-n})}{h} .
\end{equation}
Unfortunately, this approximation satisfies the identity \eqref{uniform_tens} only in the two-dimensional 
case. For the three-dimensional case, we propose the following procedure.
As suggested in \citet{visbal_02, sjogreen2014high}, one can rewrite equation \eqref{exact_tensor}  as
\begin{equation}
\displaystyle
J \nabla \xi_p = \frac{\varepsilon_{pqr}}{2}
 \frac{\partial \vecx}{\partial \xi_q} \times \frac{\partial \vecx}{\partial \xi_r}
= \frac{\varepsilon_{pqr}}{4}
\left[
\frac{\partial \phantom{ \xi_r}}{\partial \xi_r}
\left(
\frac{\partial \vecx}{\partial \xi_q} \times \vecx
\right)
 -
\frac{\partial \phantom{ \xi_q}}{\partial \xi_q}
\left(
\frac{\partial \vecx}{\partial \xi_r} \times \vecx
\right)
\right] .
\label{exact_tensor__mod}
\end{equation}
For instance, for $p=1$, one has the following expression for $\xi_1=\xi$
\begin{equation}
J \nabla \xi  = \frac{1}{2} \left[
                  \frac{\partial \phantom{\zeta}}{\partial \zeta} 
                  \left( \frac{\partial \vecx}{\partial \eta} \times \vecx \right) -
                  \frac{\partial \phantom{\eta}}{\partial \eta} 
                  \left( \frac{\partial \vecx}{\partial \zeta} \times \vecx \right) \right] .
\label{exact_tensor_1}
\end{equation}
The uniform-flow preservation property in \cite{visbal_02, sjogreen2014high} extends to any order of accuracy 
for centered schemes if the expressions in \eqref{exact_tensor_1} are discretized with 
the finite-difference approximation \eqref{fin_diff}:
\begin{equation}
\begin{array}{r@{\hspace{0mm}}l}
\displaystyle
\left. J \nabla \xi \right|_{i,j,k} = \sum_{m=1}^L \sum_{n=1}^L
                                        \frac{a_m a_n}{2 h^2} \left\{
                                        \left(  \vecx_{i,j+m,k+n}-\vecx_{i,j-m,k+n} 
                                        \right)
                                         \right.  & \times \, \vecx_{i,j,k+n}
\\*[3mm]  
\displaystyle
                                      - \left(  \vecx_{i,j+m,k-n}-\vecx_{i,j-m,k-n}
                                        \right) &
                                        \times \, \vecx_{i,j,k-n} 
\\*[3mm]  
\displaystyle
                                      - \left(  \vecx_{i,j+m,k+n}-\vecx_{i,j+m,k-n} 
                                        \right) & \times \, \vecx_{i,j+m,k}
\\*[3mm]  
\displaystyle
                                      + \left(  \vecx_{i,j-m,k+n}-\vecx_{i,j-m,k-n}
                                        \right) &
                                        \left.  \times \, \vecx_{i,j-m,k} \right\}
\end{array} .
\label{fd_tensor_1}
\end{equation}
Introducing the notation
\begin{equation}
\begin{array}{lclcl}
 \vecA_{i,j,k} & = & \left( \vecx_{i+l,j,k}-\vecx_{i-l,j,k} \right) & \times & \vecx_{i,j,k},
\\
 \vecB_{i,j,k} & = & \left( \vecx_{i,j+m,k}-\vecx_{i,j-m,k} \right) & \times & \vecx_{i,j,k},
\\
 \vecC_{i,j,k} & = & \left( \vecx_{i,j,k-n}-\vecx_{i,j,k-n} \right) & \times & \vecx_{i,j,k},
\end{array}
\end{equation}
one can rewrite \eqref{fd_tensor_1} as
\begin{equation}
\left. J \nabla \xi \right|_{i,j,k} = \sum_{m=1}^L \sum_{n=1}^L
                                        \frac{a_m a_n}{2 h^2} \left\{ 
                                        \vecB_{i,j,k+n} - \vecB_{i,j,k-n} 
                                       -\vecC_{i,j+m,k} + \vecC_{i,j-m,k} 
                                        \right\} .
\label{fd_tensor_1a}
\end{equation}
Furthermore, the first term in \eqref{discr_uf} becomes 
\begin{equation}
\begin{array}{r@{\hspace{0mm}}l}
\displaystyle
\left. \frac{\partial J \nabla \xi }{\partial \xi} \right|_{i,j,k} = & 
\displaystyle
\sum_{l=1}^L \sum_{m=1}^L \sum_{n=1}^L
                                        \frac{a_l a_m a_n}{h^3} \left\{  \right. 
\\*[5mm]  
                                      &        \vecB_{i+l,j,k+n} - \vecB_{i+l,j,k-n} 
                                              -\vecC_{i+l,j+m,k} + \vecC_{i+l,j-m,k} 
\\*[3mm]  
                                    - & \left. \vecB_{i-l,j,k+n} + \vecB_{i-l,j,k-n} 
                                             + \vecC_{i-l,j+m,k} - \vecC_{i-l,j-m,k}   \right\} .
\end{array}
\label{diff_tensor_1}
\end{equation}
We can write the other two terms in equation (\ref{exact_tensor}) with $p=2$ and $3$ in a similar way 
by index rotation
\begin{equation}
J \nabla \eta  = \frac{1}{2} \left[
                  \frac{\partial \phantom{\xi}}{\partial \xi} 
                  \left( \frac{\partial \vecx}{\partial \zeta} \times \vecx \right) -
                  \frac{\partial \phantom{\zeta}}{\partial \zeta} 
                  \left( \frac{\partial \vecx}{\partial \xi} \times \vecx \right) \right]
\label{exact_tensor_2}
\end{equation}
\begin{equation}
J \nabla \zeta  = \frac{1}{2} \left[
                  \frac{\partial \phantom{\eta}}{\partial \eta} 
                  \left( \frac{\partial \vecx}{\partial \xi} \times \vecx \right) -
                  \frac{\partial \phantom{\xi}}{\partial \xi} 
                  \left( \frac{\partial \vecx}{\partial \eta} \times \vecx \right) \right]
\label{exact_tensor_3}
\end{equation}
and approximate them by the same finite difference formulas as
\begin{equation}
\left. J \nabla \eta \right|_{i,j,k} = \sum_{l=1}^L \sum_{n=1}^L
                                        \frac{a_l a_n}{2 h^2} \left\{ 
                                        \vecC_{i+l,j,k} - \vecC_{i-l,j,k} 
                                       -\vecA_{i,j,k+n} + \vecA_{i,j,k-n} 
                                        \right\}
\label{fd_tensor_2a}
\end{equation}
and
\begin{equation}
\left. J \nabla \zeta \right|_{i,j,k} = \sum_{l=1}^L \sum_{m=1}^L
                                        \frac{a_l a_m}{2 h^2} \left\{ 
                                        \vecA_{i,j+m,k} - \vecA_{i,j-m,k} 
                                       -\vecB_{i+l,j,k} + \vecB_{i-l,j,k} 
                                        \right\}
\label{fd_tensor_3a}
\end{equation}

Then, we can rewrite the second and third terms in \eqref{discr_uf} as
\begin{equation}
\begin{array}{r@{\hspace{0mm}}l}
\displaystyle
\left. \frac{\partial J \nabla \eta }{\partial \eta} \right|_{i,j,k} = & 
\displaystyle
\sum_{l=1}^L \sum_{m=1}^L \sum_{n=1}^L
                                        \frac{a_l a_m a_n}{h^3} \left\{  \right. 
\\*[5mm]                             
                                      & \vecC_{i+l,j+m,k} - \vecC_{i-l,j+m,k} 
                                       -\vecA_{i,j+m,k+n} + \vecA_{i,j+m,k-n}
\\*[3mm]  
                                    - & \left. \vecC_{i+l,j-m,k} + \vecC_{i-l,j-m,k} 
                                       +\vecA_{i,j-m,k+n} - \vecA_{i,j-m,k-n} \right\} .
\end{array}
\label{diff_tensor_2}
\end{equation}
and
\begin{equation}
\begin{array}{r@{\hspace{0mm}}l}
\displaystyle
\left. \frac{\partial J \nabla \zeta }{\partial \zeta} \right|_{i,j,k} = & 
\displaystyle
\sum_{l=1}^L \sum_{m=1}^L \sum_{n=1}^L
                                        \frac{a_l a_m a_n}{h^3} \left\{  \right. 
\\*[5mm]                             
                                      & \vecA_{i,j+m,k+n} - \vecA_{i,j-m,k+n} 
                                       -\vecB_{i+l,j,k+n} + \vecB_{i-l,j,k+n} 
\\*[3mm]  
                                    - & \left.\vecA_{i,j+m,k-n} + \vecA_{i,j-m,k-n} 
                                       +\vecB_{i+l,j,k-n} - \vecB_{i-l,j,k-n} 
                                       \right\} .
\end{array}
\label{diff_tensor_3}
\end{equation}

If we sum all the above approximations in equation (\ref{discr_uf}), we finally obtain
\begin{equation}
\begin{array}{r@{\hspace{3mm}}l}
& \displaystyle
 \left[ \frac{\partial J \nabla \xi }{\partial \xi}                 +
       \frac{\partial J \nabla \eta }{\partial \eta}                 +
       \frac{\partial J \nabla \zeta }{\partial \zeta} \right]_{i,j,k} =  
\sum_{l=1}^L \sum_{m=1}^L \sum_{n=1}^L \frac{a_l a_m a_n}{h^3} \left\{  
    \phantom{\vecB_{i+l,j,k+n}}
    \right. 
\\*[5mm]  
                                      &        \vecB_{i+l,j,k+n} - \vecB_{i+l,j,k-n} 
                                              -\vecC_{i+l,j+m,k} + \vecC_{i+l,j-m,k} 
\\*[5mm]  
                                    - &        \vecB_{i-l,j,k+n} + \vecB_{i-l,j,k-n} 
                                             + \vecC_{i-l,j+m,k} - \vecC_{i-l,j-m,k} 
\\*[5mm]  
                                    + &        \vecC_{i+l,j+m,k} - \vecC_{i-l,j+m,k} 
                                              -\vecA_{i,j+m,k+n} + \vecA_{i,j+m,k+n} 
\\*[5mm]                                    
                                    - &        \vecC_{i+l,j-m,k} + \vecC_{i-l,j-m,k} 
                                             + \vecA_{i,j-m,k+n} - \vecA_{i,j-m,k+n} 
\\*[5mm]  
                                    + &        \vecA_{i,j+m,k+n} - \vecA_{i,j-m,k+n} 
                                              -\vecB_{i+l,j,k+n} + \vecB_{i-l,j,k+n} 
\\*[5mm]                                    
                                    - & \left. \vecA_{i,j+m,k-n} + \vecA_{i,j-m,k-n} 
                                              +\vecB_{i+l,j,k-n} - \vecB_{i-l,j,k-n}  \right\}
\\*[5mm]                                    
                                    \equiv & \quad 0
\end{array}
\label{diff_tensor_all}
\end{equation}

Therefore, if we adopt the approximation in equations (\ref{fd_tensor_1a}), (\ref{fd_tensor_2a}) and (\ref{fd_tensor_3a}), we get the correct accuracy for the metric tensor and strictly preserve uniform flows.

\section{Computation of the metric tensor on Chimera grids}\label{sec:metric_on_chimera}

The computation of the metric tensor at block boundaries with Chimera grids requires care if we want to retain the uniform flow preservation property.

To this purpose, we add $L$ ghost cell layers on each face of each actual block. Then,
on abutting block faces (with a one-to-one node matching), we simply
copy the actual positions of the nodes on the adjacent block. Conversely, on natural boundaries and 
Chimera faces (i.e. the block faces that do not match an adjacent block face), we make a linear extrapolation
of the position vector. For instance, if the ghost points $(i,j,k), \quad i=0,-1,\ldots,1-L$ are beyond a natural boundary or a Chimera face, and therefore $(1,j,k)$, $(2,j,k)$ are the two closest points to the boundary, we set
\begin{equation}
\vecx_{i,j,k} = \vecx_{1,j,k} - (1-i)  (\vecx_{2,j,k}-\vecx_{1,j,k}) \qquad i=0,-1,\ldots,1-L
\end{equation}
This procedure ensures that, even on highly curved concave boundaries where the grid lines are likely to intersect,
equation (\ref{discr_uf}) is verified. Here, of course, we must check that the sign on the Jacobian is the same  $s_J$ as for internal points: if this condition is not verified, we enforce the correct sign as $ J_{i,j,k} = s_J |J_{i,j,k}|$.



%\Urlmuskip=0mu plus 1mu\relax
 \bibliographystyle{model1-num-names}
\bibliography{references.bib}
%\bibliographystyle{plain}






% \end{thebibliography}

\end{document}